\documentclass{article}

\usepackage[main,final]{neurips_2026}  
\usepackage[utf8]{inputenc}
\usepackage[T1]{fontenc}
\usepackage{hyperref}
\usepackage{url}
\usepackage{booktabs}
\usepackage{amsfonts}
\usepackage{amsmath}
\usepackage{amssymb}
\usepackage{nicefrac}
\usepackage{microtype}
\usepackage[table]{xcolor}
\usepackage[final]{graphicx}
\usepackage{multirow}
\usepackage{array}
\usepackage{subcaption}

\usepackage{tikz}
\usetikzlibrary{positioning,arrows.meta,calc,fit}
\usepackage{amsthm}
\theoremstyle{plain}
\newtheorem{theorem}{Theorem}
\newtheorem{lemma}{Lemma}
\newtheorem{proposition}{Proposition}
\theoremstyle{definition}
\newtheorem{definition}{Definition}

\newcommand{\milddrop}[1]{\textcolor{black!55}{#1}}
\newcommand{\drop}[1]{\textcolor{black!30}{#1}}

\newcommand{\zold}{\mathbf{z}_{\mathrm{old}}}
\newcommand{\znew}{\mathbf{z}_{\mathrm{new}}}
\newcommand{\zfinal}{\mathbf{z}_{\mathrm{final}}}
\newcommand{\Lorth}{\mathcal{L}_{\mathrm{orth}}}
\newcommand{\lambdaorth}{\lambda_{\mathrm{orth}}}

\title{You CAN Teach an Old Model New Tricks: Domain Adaptation via Complementary Subspace Expansion}

\author{
  Donghoon Han \enspace
  SungHyun Moon \enspace
  SeungJae Lee \\
  Dnotitia Inc.\\
  \texttt{\{dhk1349,\,sunghyun,\,seungjae.lee\}@dnotitia.com}
}

\begin{document}

\maketitle

\begin{abstract}
Adapting a deployed retrieval encoder can improve domain quality while degrading general retrieval and making existing corpus embeddings incompatible. We introduce \textbf{Complementary Subspace Expansion (CoSE)}, which preserves the frozen encoder's output and appends a complementary channel trained with both blended and standalone retrieval losses. An inference-time blending weight controls the domain--general trade-off, with a frozen endpoint that preserves every legacy score. Across four domains and three seeds, CoSE trains $6.4$M parameters and keeps all reported general benchmark means across three training seeds within $4.8$ percentage points of Frozen at equal channel weighting, while validation-selected weights improve domain retrieval over Frozen. Complementary-channel truncation and partial branching provide storage and encoding-cost trade-offs. Serving benchmarks and comparisons with two-table alternatives characterize the deployment trade-off: CoSE retains an auditable legacy-score endpoint within one expanded representation, at the cost of additional compute and corpus backfill.
\end{abstract}

\section{Introduction}
\label{sec:intro}

Vision-language embedding models underpin large-scale multimodal retrieval, where queries and documents are mapped to a shared vector space and compared against an index built from massive corpora. In deployment, this index is often the largest asset in the serving stack: re-embedding a billion-document corpus can take weeks of GPU time. Adapting a deployed model to a new domain therefore incurs a second cost beyond fine-tuning: index migration.

Current adaptation methods force a trade-off. PEFT methods such as LoRA~\cite{LoRALowRankAdaptation} and DoRA~\cite{DoRAWeightDecomposedLowRankAdaptation} modify the base encoder's weights, invalidating old indices and often degrading general retrieval. BCT~\cite{TowardsBackwardCompatibleRepresentationLearning,BCAlignerBackwardCompatibleAlignment,AdvBCTAdversarialBackwardCompatibleTraining} relaxes compatibility into an approximate distillation objective. L2C~\cite{L2CLearningToComplement}, the closest architectural neighbor, combines a frozen encoder with a parallel network but still changes the final feature, breaking exact index compatibility. Keeping the base frozen preserves general retrieval, but provides no domain adaptation.

We ask whether the old model can learn new tricks \emph{without being touched}---whether backward compatibility can be guaranteed by construction. We propose \textbf{Complementary Subspace Expansion (CoSE)}, a side-branch LoRA architecture in which the base encoder stays frozen while a jointly trained side branch produces a complementary embedding $\znew$. The final retrieval embedding is the concatenation $\zfinal = [\zold \,\|\, \znew]$, where $\zold$ is structurally preserved from the frozen encoder's output on the same input. At the frozen endpoint, CoSE preserves every query--document score against the legacy corpus. We call this exact backward compatibility (BC); using the complementary channel requires additional corpus embeddings. The complementary channel $\znew$ adds domain signal through contrastive training, while a blending scalar $\alpha$ shifts weight between $\zold$ and $\znew$ at inference without retraining.

\begin{figure}[t]
\centering
\includegraphics[width=0.95\linewidth]{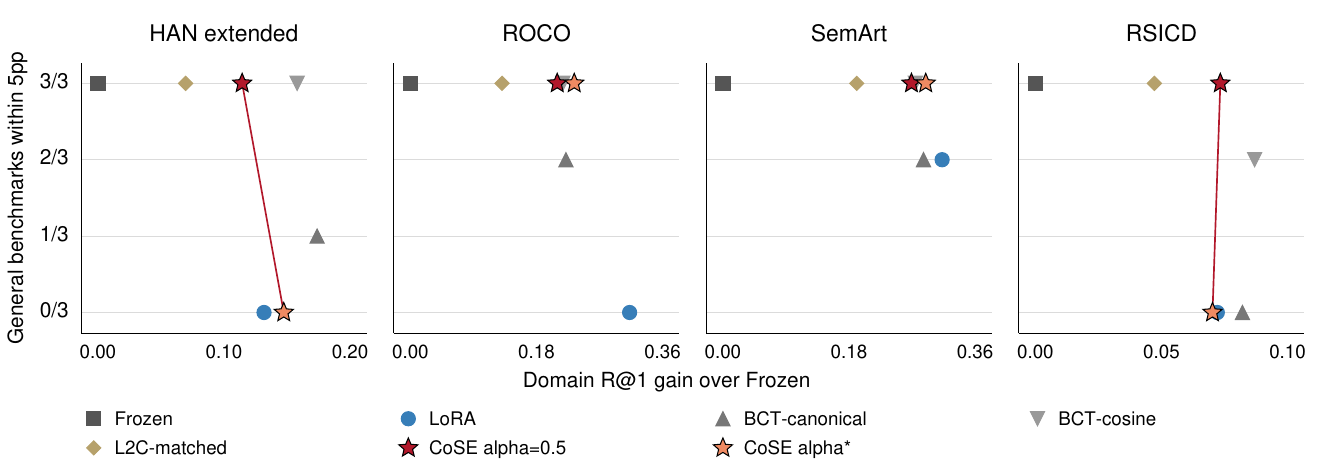}
\caption{\textbf{Domain quality and general preservation after full corpus migration.} $x$: domain R@$1$ gain over Frozen, avg(i2t,t2i); $y$: number of general benchmarks within $5$pp of Frozen. Points use Table~\ref{tab:main} results and each method's populated corpus representation; SemArt/RSICD LoRA general coverage uses seed $0$, while its domain gain averages three seeds. The red segment connects CoSE's fixed and validation-selected blends from the same checkpoint family; it is not a measured intermediate trajectory.}
\label{fig:pareto-preview}
\end{figure}

Figure~\ref{fig:pareto-preview} summarizes fully migrated domain quality and general-preservation coverage under a $5$pp budget. CoSE's fixed and validation-selected blends expose different operating points; the experiments separately evaluate legacy-score preservation, deployment alternatives, and serving cost.

\paragraph{Contributions.}
\begin{itemize}
\item \textbf{Domain adaptation with an unchanged legacy channel.}
CoSE preserves the frozen encoder's output as a coordinate block and learns a complementary retrieval channel. The frozen endpoint preserves legacy scores, while blended retrieval adds domain signal after corpus backfill.
\item \textbf{Joint training with controllable deployment costs.}
Contrastive supervision of both the blended and complementary representations supports different inference-time operating points. Optional Matryoshka supervision on HAN and partial branching provide storage and encoding-cost trade-offs.
\item \textbf{Evaluation of quality, preservation, and deployment cost.}
Four-domain experiments quantify domain retrieval and general retention, while serving benchmarks and two-table comparisons establish the benefits and limits of preserving the legacy channel within an expanded representation.
\end{itemize}

\paragraph{Deployment scope.}
CoSE does not eliminate corpus backfill: domain gains require computing the complementary channel for indexed documents. We recommend serving the frozen endpoint until backfill completes, then switching to the selected blend. After migration, the frozen channel remains available and the blending weight controls the domain--general trade-off without retraining.

\section{Related Work}
\label{sec:related}

\paragraph{Backward-compatible representation learning.}
BCT~\cite{TowardsBackwardCompatibleRepresentationLearning} and variants (adversarial~\cite{AdvBCTAdversarialBackwardCompatibleTraining}, aligner-based~\cite{BCAlignerBackwardCompatibleAlignment}, forward-compatible~\cite{ForwardCompatibleTrainingApple}, efficient-update~\cite{FastFillBackwardCompatibleFeatureFilling}) treat compatibility as a distillation target. CoSE instead preserves the base's outputs \emph{structurally} as the first
$d$ coordinates of $F_\theta=[\zold;\znew]$ by expanding the score channel rather than perturbing within it (Sec.~\ref{sec:method_problem}).

\paragraph{PEFT with weight-space constraints.}
LoRA~\cite{LoRALowRankAdaptation} and DoRA~\cite{DoRAWeightDecomposedLowRankAdaptation} provide parameter-efficient weight adaptation, while OPLoRA~\cite{OPLoRAOrthogonalLowRankAdaptation} and MiLoRA~\cite{MiLoRAMinorSubspaceLoRA} constrain the update subspace. These methods do not enforce exact equality with the frozen encoder's outputs. CoSE's complementarity is analyzed in \emph{output} space between $\zold$ and $\znew$; it does not require an explicit orthogonality constraint.

\paragraph{Frozen-backbone complementary-network architectures --- closest comparable.}
Side-Tuning~\cite{SideTuningTransferLearning} and LST~\cite{LadderSideTuning} established the frozen-backbone-plus-side pattern. L2C~\cite{L2CLearningToComplement} is closest: a frozen CLIP~\cite{CLIPLearningTransferableVisualModels} encoder plus a parallel network combined via ``reverted attention,'' $\tilde{\mathbf{I}}(x) = \mathbf{I}(x) + \mathrm{CP}^{\mathrm{RT}}(x)$. CoSE differs on three axes: (i) L2C's additive form modifies the final feature, breaking BC; CoSE's concatenation keeps $\zold$ verbatim; (ii) L2C is classification-focused, CoSE is retrieval-formulated; (iii) CoSE uses in-backbone LoRA, enabling partial branching. We compare against parameter-matched and full-capacity retrieval-adapted L2C in Sec.~\ref{sec:domain}.

\paragraph{Multimodal embedding adaptation.}
CLIP-family models~\cite{CLIPLearningTransferableVisualModels,SigLIP2ImprovedSigmoidLossForLanguageImage,MetaCLIP2Demystifying}, domain-specialized descendants~\cite{BiomedCLIPForBiomedicalImages,RemoteCLIPForRemoteSensing}, and VLM-derived embedders~\cite{VLM2VecUniversalMultimodalEmbeddings,MMEmbedUniversalMultimodalRetrieval,GMEUniversalMultimodalRetrieval} adapt via full or projector fine-tuning; none target BC. CoSE's side-branch recipe is applicable on top of any such base.

\paragraph{CLIP adapter family.}
CLIP-Adapter~\cite{CLIPAdapterBetterVisionLanguageModelsWithFeatureAdapters} and Tip-Adapter~\cite{TipAdapterTrainingFreeCLIPAdapter} add small trainable modules on a frozen CLIP for few-shot classification. They share CoSE's frozen-backbone premise but are designed for classification-time feature/logit adaptation rather than deployed retrieval embedding expansion.

\paragraph{Continual learning, Matryoshka, and InfoNCE.}
Forgetting mitigation via parameter regularization (EWC~\cite{EWCElasticWeightConsolidation}) or output distillation (LwF~\cite{LwFLearningWithoutForgetting}) reduces but does not eliminate drift; CoSE achieves exact base preservation architecturally. We adopt Matryoshka Representation Learning~\cite{MatryoshkaRepresentationLearning} for $\znew$ truncation; partial branching is algorithmically similar to top-layer adapters~\cite{AdapterHoulsby2019} but motivated by serving cost. Contrastive training uses InfoNCE~\cite{InfoNCEOordCPC} with in-batch negatives; we use no momentum queue~\cite{MoCoMomentumContrast}; small-batch quality remains a limitation (App.~\ref{app:batch}).

\section{Methodology}
\label{sec:method}

\subsection{Problem Formulation: Exact Compatibility as a Score Constraint}
\label{sec:method_problem}

Let $f_0:\mathcal{X}\to\mathbb{R}^d$ denote a frozen vision-language embedding encoder and
\[
    \zold(x)=\frac{f_0(x)}{\|f_0(x)\|_2}
\]
the L2-normalized legacy embedding. A deployed corpus
$\mathcal{D}=\{d_i\}_{i=1}^N$ has the fixed index
$\mathcal{I}_0=\{\zold(d_i)\}_{i=1}^N$. Given paired domain data
$\mathcal{T}_{\mathrm{dom}}$, the target is domain-adapted retrieval with
(i) an exactly auditable frozen endpoint, (ii) improved domain quality, and
(iii) general-retrieval preservation. CoSE satisfies the first requirement by
expanding the score channel instead of perturbing the deployed one.

\begin{definition}[Exact backward compatibility of legacy scores]
\label{def:legacy_score_compat}
Let $Z_{\mathrm{old}}\in\mathbb{R}^{N\times d}$ have rows $\zold(d_i)^\top$. A query encoder $g:\mathcal{X}\to\mathbb{R}^d$ preserves every legacy score if
\begin{equation}
Z_{\mathrm{old}}g(q)=Z_{\mathrm{old}}\zold(q)\quad\text{for every }q.
\end{equation}
\end{definition}

If the legacy document matrix has full column rank, preserving every legacy score within the original embedding channel forces the adapted query embedding to equal the frozen embedding: $Z_{\mathrm{old}}(g(q)-\zold(q))=0$ implies $g(q)=\zold(q)$. CoSE therefore retains the legacy channel and learns an additional one. This observation concerns exact score equality, which is stronger than preserving rankings.

\subsection{Complementary Expansion and Architecture}
\label{sec:method_cose}

CoSE represents each input by
\[
    F_\theta(x)=
    \begin{bmatrix}
    \zold(x)\\
    \znew(x;\theta)
    \end{bmatrix}
    \in\mathbb{R}^{d+m},
\]
where $\znew(x;\theta)\in\mathbb{R}^m$ is trainable and domain-specialized.

If a fixed full-rank linear readout recovers the frozen embedding from an expanded representation, an invertible coordinate transformation expresses that representation as a frozen block plus a complementary block. CoSE implements this form directly, making the legacy readout a coordinate slice. Formal statements and proofs are in Appendix~\ref{app:compatibility-theory}.

\label{sec:method_architecture}

\begin{figure}[t]
  \centering
  \includegraphics[width=0.55\linewidth]{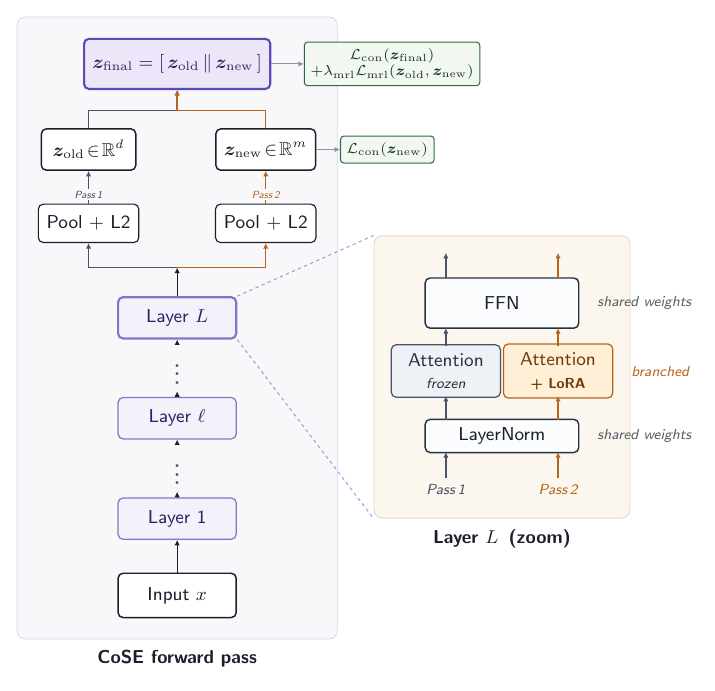}
  \caption{CoSE architecture. Pass~1 runs the frozen encoder and produces the
  preserved legacy embedding $\zold$. Pass~2 is a trainable side-branch LoRA
  pathway that produces the
  complementary embedding $\znew$. The final representation is
  $F_\theta=[\zold\,;\,\znew]$. Shared weights act separately on each pathway; MRL supervises truncated blends and is used only on HAN in the main experiments.}
  \label{fig:architecture}
\end{figure}

The reference implementation uses one forward over a duplicated batch $[x;x]$. The backbone weights are frozen, and hooks on attention projections add LoRA deltas only to the second half. Each half computes its own hidden states from the input; the complementary half does not consume the first half's intermediate states. The first half produces $\zold$ and the second produces $\znew$, each with last-token pooling and L2 normalization. The pooled $\zold$ is detached in training. Thus the first half is independent of the LoRA parameters.

The default width is $m=d$. For Qwen3-VL-Embedding-2B ($L=28$, $d=2048$), rank-$16$ LoRA on $Q/K/V/O$ adds $6.4$M trainable parameters ($0.30\%$ of the base). Complementary coordinates can be truncated and renormalized at deployment. An optional Matryoshka loss supports this use on HAN; it is not used by the main-table ROCO, SemArt, or RSICD checkpoints.

\begin{lemma}[Index preservation by construction]
\label{lem:index_preservation}
For every input $x$ and parameter value $\theta$, $F_\theta(x)_{1:d}=\zold(x)$. Therefore, for any query $q$ and legacy document $d_i$,
\[
    \langle F_\theta(q)_{1:d},\zold(d_i)\rangle
    =
    \langle \zold(q),\zold(d_i)\rangle,
\]
which is exactly the frozen-model score. The operating point using only the
first $d$ coordinates recovers the frozen ranking over the legacy index.
\end{lemma}

\begin{proof}
The first pathway uses frozen weights and contains no trainable operation from
the side branch. The stop-gradient makes this separation explicit in the
computation graph. Hence the first $d$ coordinates are independent of
$\theta$ and equal $\zold(x)$ up to kernel-level floating-point determinism.
\end{proof}

At the frozen endpoint,
the legacy index remains valid without rewriting. For adapted retrieval with
nonzero contribution from $\znew$, corpus documents additionally store
complementary coordinates, while the legacy $\zold$ entries remain unchanged. The guarantee concerns underlying similarity scores. ANN indices built over different representations may return different candidates; retaining the unchanged legacy ANN index preserves the original search path for rollback.

\subsection{Retrieval Scores and Operating Points}
\label{sec:method_alpha_kernel}
\label{sec:method_alpha_selection}

For a query-document pair $(q,d)$, define
\[
    r_0(q,d)=\langle \zold(q),\zold(d)\rangle,
    \qquad
    r_1(q,d)=\langle \znew(q),\znew(d)\rangle .
\]
CoSE uses
\begin{equation}
    s_\alpha(q,d)
    =
    \alpha r_0(q,d)+(1-\alpha)r_1(q,d),
    \qquad
    \alpha\in[0,1].
    \label{eq:alpha_score}
\end{equation}

The weighted concatenation $\phi_\alpha(x)=[\sqrt{\alpha}\,\zold(x);\sqrt{1-\alpha}\,\znew(x)]$ induces $s_\alpha$ and a positive-semidefinite kernel for every $\alpha\in[0,1]$. At $\alpha=1$, it reduces to the frozen score; at $\alpha=0.5$, it equals cosine similarity of the unweighted, branch-normalized concatenation.

We report $\alpha=0.5$ as a tuning-free reference and select $\alpha^*$ on domain validation i2t R@$1$ over a fixed grid. If general retention is required, the same sweep can instead maximize domain quality subject to a specified general-degradation budget; this need not select the domain-optimal weight. Validation selection and general preservation are separate criteria: a domain-validation optimum need not meet a general-quality budget. We therefore evaluate the fixed weight directly, rather than infer its preservation from the stability of a selected weight.

\paragraph{Changing the blend with fixed stored vectors.}
The symmetric representation $\phi_\alpha$ defines the score, but need not be the storage convention for every weight. For example, keeping $D(d)=\phi_{1/2}(d)$ fixed and using
\[
 Q_\alpha(q)=[\sqrt{2}\alpha\zold(q);\sqrt{2}(1-\alpha)\znew(q)]
\]
gives $\langle Q_\alpha(q),D(d)\rangle=s_\alpha(q,d)$ without rewriting documents. This is an inner-product construction, with no further query normalization. It establishes score equivalence, not invariant approximate-search behavior: changing the query weights can change HNSW traversal, recall, and cost. Our serving-cost measurements use the fixed $\alpha=0.5$ operating point; we do not measure a query-only reweighting cost or recall sweep. A separate frozen-endpoint ANN diagnostic is described in Appendix~\ref{app:reproducibility}. Exact legacy scores at $\alpha=1$ therefore remain distinct from reproducing the legacy ANN candidate set.

\subsection{Training Objective}
\label{sec:method_objective}

For a minibatch $\mathcal{B}=\{(u_i,v_i)\}_{i=1}^B$ of ordered query--positive pairs, the reported implementation uses the directional contrastive loss
\[
 \mathcal{L}_{\mathrm{NCE}}(s)=-\frac1B\sum_{i=1}^B
 \log\frac{\exp(s(u_i,v_i)/\tau)}
 {2\exp(s(u_i,v_i)/\tau)+\sum_{j\in\mathcal{N}_i}\exp(s(u_i,v_j)/\tau)},
\]
where $\mathcal{N}_i$ excludes the diagonal and other pairs with the same image ID. The factor two reflects the positive logit and retained in-batch diagonal; the former is the target (Appendix~\ref{app:reproducibility}). Training is directional: image-to-text on ROCO, SemArt, and RSICD and the sampled HAN pair types (Appendix~\ref{app:datasets}). Evaluation reports both directions. All CoSE contrastive terms use this implemented loss. It is not standard InfoNCE up to a constant: writing the positive exponential as $P$ and the negative sum as $N$, the difference is $\log(1+P/(P+N))$, which depends on the embeddings. Baseline objectives are distinguished in Appendix~\ref{app:baselines}.

We define the contrastive training recipe as
\begin{equation}
    \mathcal{L}_{\mathrm{CoSE}}
    =
    \mathcal{L}_{\mathrm{NCE}}(s_{1/2})
    +
    \lambda_{\mathrm{new}}\mathcal{L}_{\mathrm{NCE}}(r_1)
    +
    \lambda_{\mathrm{mrl}}\mathcal{L}_{\mathrm{MRL}}(\zold,\znew).
    \label{eq:cose_loss}
\end{equation}
The main-table checkpoints minimize $\mathcal{L}_{\mathrm{CoSE}}+\lambdaorth\Lorth$ with $\lambdaorth=0.1$; the paired loss controls use $\lambdaorth=0$. Appendix~\ref{app:orthogonality} defines the penalty. The first term trains the default unweighted concatenation; the second term
anchors the standalone complementary endpoint; the $\alpha=1$ endpoint is inherited from the frozen
encoder.

For Matryoshka training, let $\mathcal{M}=\{256,512,1024,2048\}$ be the reference truncation set, with $m'\le m$, and define
\[
    \znew^{(m')}(x)=\operatorname{norm}(\znew(x)_{1:m'}),\qquad
    r_1^{(m')}(q,d)=\langle \znew^{(m')}(q),\znew^{(m')}(d)\rangle .
\]
Then
\[
    \mathcal{L}_{\mathrm{MRL}}(\zold,\znew)
    =
    \frac{1}{|\mathcal{M}|}\sum_{m'\in\mathcal{M}}
    \mathcal{L}_{\mathrm{NCE}}(\tfrac12 r_0+\tfrac12 r_1^{(m')}).
\]
The MRL term averages losses on the blended, branch-normalized concatenation $[\zold;\znew^{(m')}]$, including full width. It does not supervise $\znew$ alone. We use $\lambda_{\mathrm{new}}=1$, $\tau=0.02$, and $\lambda_{\mathrm{mrl}}=0.5$ for HAN and $0$ for ROCO/SemArt/RSICD. MRL is an optional trade-off for truncated blended deployment; its measured benefit is limited to HAN (Section~\ref{sec:loss-ablation}). Training configurations are given in Appendix~\ref{app:reproducibility}.

\subsection{Partial Branching and Serving Cost}
\label{sec:method_partial_branching}

Full CoSE runs both the frozen pathway and the complementary pathway, giving
approximately $2\times$ base forward cost and $d+m$ stored coordinates per
document. Let $K\in\{0,\ldots,L\}$ denote the number of shared prefix layers.
The frozen pathway still runs all $L$ layers and produces $\zold$ exactly, while
the complementary pathway starts from the detached hidden state after layer
$K$ and runs the remaining $L-K$ layers with LoRA adapters.
The analytical base-equivalent cost $C(K)=1+(L-K)/L$ assumes the prefix is computed once. The duplicated-batch reference recomputes it and does not realize this saving. Output-equivalent prefix sharing realizes this saving in the tested implementation (Appendix~\ref{app:partial-branching-results}).
Since the frozen pathway is unchanged for every $K$,
Lemma~\ref{lem:index_preservation} holds regardless of branching depth.
Storage is $d+m$ at full width and $d+m'$ under Matryoshka truncation with
$m'\leq m$.

\section{Experiments}
\label{sec:experiments}

\subsection{Setup}
\label{sec:setup}

\textbf{Backbone.} Qwen3-VL-Embedding-2B~\cite{Qwen3VLEmbedding} ($L{=}28$, $d{=}2048$, last-token pool). Additional experiments include an 8B diagnostic, a 2B batch-size control, and transfer to SigLIP2-so400m (\S\ref{sec:transfer}).
\textbf{Datasets.} Four domains spanning broadcast/cultural, medical, art, and satellite imagery: HAN extended\footnote{Extended HAN~\cite{NIACulturalDataset2025}; base on AIHub (\url{https://aihub.or.kr/aihubdata/data/view.do?dataSetSn=71866}); extended captions release pending review.}, ROCO~\cite{ROCODataset}, SemArt~\cite{SemArtDataset}, RSICD~\cite{RSICDDataset}. General-retrieval eval: Flickr30K~\cite{Flickr30KDataset}, COCO 2017 val~\cite{COCODataset} ($1000$ pairs each), ImageNet-1k~\cite{ImageNetDataset} zero-shot ($5000$ images), all seed $42$. Per-domain split sizes and protocol in Appendix~\ref{app:datasets}.
\textbf{Training.} The CoSE checkpoints use batch $64$ with MRL weight $0.5$ for HAN, and batch $32$ without MRL for ROCO/SemArt/RSICD. They use lr $2\times10^{-4}$, $200$-step warmup and cosine decay over $8000$ steps, $\tau=0.02$, and rank-$16$ LoRA on all $Q/K/V/O$ projections ($6.42$M trainable). The main-table checkpoints include $\lambdaorth=0.1$; paired ablations use zero. LoRA and BCT use the same rank and reference schedule; actual batch settings are distinguished in Appendix~\ref{app:baselines}. L2C has parameter-matched and full-capacity variants.

\textbf{Evaluation.} Domain retrieval uses the same $1000$ sampled image records across methods (sampling seed $0$), with one selected caption and one matching positive per record. Caption rules are in Appendix~\ref{app:datasets}. Fully migrated retrieval uses each method's populated corpus representation; cross retrieval against an unchanged frozen corpus is evaluated separately. R@1 in the main text; R@5/R@10 and per-direction i2t/t2i breakdowns are reported in Appendix~\ref{app:preservation-detail}. CoSE $\alpha$ swept on val over 12 values; $\alpha^{\!*}=\arg\max_\alpha \mathrm{val\text{-}i2t\text{-}R}@1$ (ties choose the smallest $\alpha$); test reports both $\alpha{=}0.5$ and $\alpha^{\!*}$.
\textbf{Baselines.} \emph{Frozen} (unadapted base), \emph{LoRA} (standard PEFT, matched rank), \emph{BCT-canonical}~\cite{TowardsBackwardCompatibleRepresentationLearning} (symmetric cross-encoded InfoNCE), \emph{BCT-cosine} (feature-alignment distillation), and \emph{L2C}~\cite{L2CLearningToComplement} (reverted attention, retrieval-adapted). We additionally evaluate a full-capacity retrieval-adapted L2C, a Frozen+LoRA score blend, and a two-table system retaining separate Frozen and LoRA embeddings. Full definitions in Appendix~\ref{app:baselines}.

\subsection{Main Domain Performance}
\label{sec:domain}

Table~\ref{tab:main} reports test-split R@1 on four domains.

\begin{table}[t]
\centering
\caption{\textbf{Fully migrated retrieval and general preservation.} Every method uses its corresponding populated corpus representation. Domain and COCO/Flickr values are avg(i2t,t2i) R@$1$; ImageNet is zero-shot Top-$1$. Domain values are mean$\pm$sample std across three training seeds (0/42/123). General evaluation uses shared samples across methods ($1000$ retrieval pairs, $5000$ ImageNet images). General values are three-seed means except SemArt/RSICD LoRA, reported for seed $0$ only; their domain values remain three-seed means. The \emph{endpoint} mark for CoSE means exact legacy scores are available at $\alpha=1$, not that scores at the displayed $\alpha$ equal Frozen; $\approx$ denotes approximate compatibility training for single-encoder baselines. Grayscale marks degradation from Frozen: at most $3$pp, $3$--$10$pp, and over $10$pp. CoSE $\alpha^*$ is selected per seed on validation i2t R@$1$.}
\label{tab:main}
\scriptsize
\setlength{\tabcolsep}{3pt}
\begin{tabular}{l l c | c | ccc}
\toprule
Trained on & Method & \shortstack{Legacy-score\\endpoint} & \shortstack{Domain R@1\\avg(i2t,t2i)} & \shortstack{INet\\Top-1} & \shortstack{COCO\\avg} & \shortstack{Flickr\\avg} \\
\midrule
\emph{Frozen} & (no adaptation) & \textcolor{blue}{$\checkmark$} & 0.320 / 0.142 / 0.204 / 0.088 & 0.644 & 0.687 & 0.880 \\
\midrule
\multirow{6}{*}{HAN extended} & LoRA              & \textcolor{red}{$\times$} & $0.450 \pm 0.007$              & \drop{0.524} & \drop{0.507} & \drop{0.776} \\
                              & BCT-canonical     & $\approx$                 & \textbf{$0.492 \pm 0.001$}     & \drop{0.542} & \drop{0.571} & 0.858 \\
                              & BCT-cosine        & $\approx$                 & $0.476 \pm 0.003$              & 0.623 & \milddrop{0.647} & 0.888 \\
                              & L2C-matched       & \textcolor{red}{$\times$} & $0.388 \pm 0.005$              & 0.625 & 0.682 & 0.865 \\
\rowcolor{gray!8}             & \textbf{CoSE $\alpha{=}0.5$} & \textcolor{blue}{$\checkmark$} & $0.433\pm0.006$ & 0.635 & 0.683 & 0.881 \\
\rowcolor{gray!15}            & \textbf{CoSE $\alpha^{\!*}\approx0.12$} & \textcolor{blue}{$\checkmark$} & $0.465\pm0.005$ & \drop{0.531} & \drop{0.507} & \drop{0.757} \\
\midrule
\multirow{6}{*}{ROCO}         & LoRA              & \textcolor{red}{$\times$} & \textbf{$0.457 \pm 0.007$}     & \drop{0.491} & \drop{0.568} & \drop{0.767} \\
                              & BCT-canonical     & $\approx$                 & $0.365 \pm 0.003$              & \milddrop{0.593} & \milddrop{0.645} & \milddrop{0.844} \\
                              & BCT-cosine        & $\approx$                 & $0.361 \pm 0.001$              & 0.622 & 0.660 & 0.857 \\
                              & L2C-matched       & \textcolor{red}{$\times$} & $0.273 \pm 0.007$              & 0.637 & 0.683 & 0.883 \\
\rowcolor{gray!8}             & \textbf{CoSE $\alpha{=}0.5$} & \textcolor{blue}{$\checkmark$} & $0.353\pm0.002$ & 0.638 & 0.684 & 0.893 \\
\rowcolor{gray!15}            & \textbf{CoSE $\alpha^{\!*}\approx0.25$} & \textcolor{blue}{$\checkmark$} & $0.377\pm0.007$ & \milddrop{0.600} & \milddrop{0.655} & 0.865 \\
\midrule
\multirow{6}{*}{SemArt}       & LoRA              & \textcolor{red}{$\times$} & \textbf{$0.513 \pm 0.003$}     & 0.622 & \milddrop{0.615} & \milddrop{0.850} \\
                              & BCT-canonical     & $\approx$                 & $0.487 \pm 0.001$              & \milddrop{0.602} & \milddrop{0.624} & \milddrop{0.844} \\
                              & BCT-cosine        & $\approx$                 & $0.476 \pm 0.007$              & 0.630 & 0.669 & 0.869 \\
                              & L2C-matched       & \textcolor{red}{$\times$} & $0.393 \pm 0.003$              & 0.633 & 0.669 & 0.866 \\
\rowcolor{gray!8}             & \textbf{CoSE $\alpha{=}0.5$} & \textcolor{blue}{$\checkmark$} & $0.470\pm0.004$ & 0.650 & 0.685 & 0.889 \\
\rowcolor{gray!15}            & \textbf{CoSE $\alpha^{\!*}\approx0.30$} & \textcolor{blue}{$\checkmark$} & $0.490\pm0.007$ & 0.629 & \milddrop{0.656} & 0.870 \\
\midrule
\multirow{6}{*}{RSICD}        & LoRA              & \textcolor{red}{$\times$} & $0.157 \pm 0.011$              & \drop{0.287} & \drop{0.453} & \drop{0.645} \\
                              & BCT-canonical     & $\approx$                 & $0.167 \pm 0.011$              & \drop{0.449} & \drop{0.567} & \drop{0.763} \\
                              & BCT-cosine        & $\approx$                 & \textbf{$0.172 \pm 0.006$}     & \drop{0.511} & \milddrop{0.637} & \milddrop{0.831} \\
                              & L2C-matched       & \textcolor{red}{$\times$} & $0.133 \pm 0.005$              & 0.631 & 0.679 & 0.876 \\
\rowcolor{gray!8}             & \textbf{CoSE $\alpha{=}0.5$} & \textcolor{blue}{$\checkmark$} & $0.159\pm0.004$ & \milddrop{0.596} & 0.675 & 0.865 \\
\rowcolor{gray!15}            & \textbf{CoSE $\alpha^{\!*}\approx0.23$} & \textcolor{blue}{$\checkmark$} & $0.156\pm0.006$ & \drop{0.497} & \drop{0.586} & \milddrop{0.789} \\
\bottomrule
\end{tabular}
\end{table}

\label{sec:general-pres}
At $\alpha=0.5$, CoSE improves domain retrieval on all four domains while all general mean metrics remain within $4.8$pp of Frozen. Domain-validation selection yields another operating point, trailing the numerical domain leader by $2.7/8.0/2.3/1.7$pp on HAN/ROCO/SemArt/RSICD. Domain-only selection exceeds the $5$pp general budget on HAN and RSICD for every seed and on ROCO for one of three seeds, although ROCO's mean satisfies it. L2C-matched also satisfies the $5$pp budget on all four domains at the level of benchmark means, and BCT-cosine on three. CoSE at $\alpha=0.5$ has higher domain R@$1$ than L2C-matched on all four domains; general retention alone is not unique to CoSE. The fixed and tuned weights therefore serve different operating objectives. Seed variation is conditional on the training protocol (Appendix~\ref{app:reproducibility}); gaps are interpreted descriptively.

\paragraph{Full-capacity L2C.}
With approximately $155$M trainable parameters, L2C-full obtains domain i2t R@$1$ of $0.358/0.453/0.142$ on ROCO/SemArt/RSICD, compared with $0.355/0.474/0.165$ for CoSE at $\alpha=0.5$ (CoSE: three-seed means; L2C-full: seed $0$ only). Its COCO avg(i2t,t2i) R@$1$ is $1.2$--$2.1$pp below Frozen; Flickr and ImageNet also remain within $5$pp. The domain i2t comparisons are separate from Table~\ref{tab:main}'s bidirectional averages.

\subsection{Deployment Alternatives and Migration}
\label{sec:deployment-comparison}
A two-table alternative retains Frozen corpus embeddings and builds a separate LoRA table. With known query domains, hard routing selects one encoder and searches its table over the full corpus. It exceeds CoSE at fixed $\alpha=0.5$ on all three tested mixed-traffic domains (Table~\ref{tab:two-table}). CoSE therefore targets deployments valuing an expanded representation and a domain--general blend without a per-query routing decision, rather than universally higher retrieval quality.

\begin{table}[t]
\centering\small
\caption{Mixed-traffic text-to-image R@$1$ (mean$\pm$sample std, three seeds) on the union of domain and Flickr documents, with domain-test and Flickr text queries. Hard routing assumes known query domains. SemArt/RSICD LoRA seeds 42/123 are retrained replicas, distinct from the main-table runs (Appendix~\ref{app:reproducibility}). This evaluation differs from the ROCO-only serving benchmark.}
\label{tab:two-table}
\begin{tabular}{lccc}\toprule
System & ROCO & SemArt & RSICD \\ \midrule
Two-table, known-domain routing & $0.672\pm0.004$ & $0.693\pm0.004$ & $0.515\pm0.003$ \\
CoSE, fixed $\alpha=0.5$ & $0.623\pm0.003$ & $0.680\pm0.003$ & $0.510\pm0.003$ \\
\bottomrule\end{tabular}
\end{table}

During partial corpus backfill, blending covered and uncovered documents creates coverage bias. At $25\%$ coverage, HAN R@$1$ falls from $0.341$ to $0.270$ and ROCO from $0.151$ to $0.134$; all four domains show excess top-$1$ selection of covered documents. We therefore serve the frozen endpoint until backfill completes, then switch atomically. Blue-green migration can provide interim continuity for other methods too; CoSE retains the frozen score channel inside the final model after migration.

\subsection{Serving Cost}
\label{sec:serving}
We benchmark HNSW on $65$k distinct ROCO documents at ANN recall@$10\ge0.95$. Table~\ref{tab:serving} reports distance computations per query (ndis), an evaluated-vector payload proxy (ndis $\times$ dimension $\times4$ bytes), and index size projected to $100$k documents. The two-table row stores both tables but reports search work on its selected LoRA path; it is not a mixed-traffic average or a matched quality--cost Pareto comparison with Table~\ref{tab:two-table}.

\begin{table}[t]
\centering\small\setlength{\tabcolsep}{5pt}
\caption{ROCO HNSW search work and projected storage, excluding any additional legacy index retained for ANN rollback (Appendix~\ref{app:deployment}). Payload is a proxy, not measured memory traffic or encoder throughput. The $2304$-dimensional CoSE row truncates only the complementary channel; full-width domain/general performance must not be attributed to this configuration.}
\label{tab:serving}
\begin{tabular}{lrrrr}\toprule
Index & Dim & ndis/query & Payload & Size ($100$k) \\ \midrule
Frozen & 2048 & 797 & 6.5 MB & 846 MB \\
LoRA & 2048 & 1049 & 8.6 MB & 846 MB \\
Two-table, hard routing & $2048$/table & 1049 & 8.6 MB & 1692 MB \\
CoSE full & 4096 & 532 & 8.7 MB & 1666 MB \\
CoSE $[2048\|256]$ & 2304 & 837 & 7.7 MB & 949 MB \\
\bottomrule\end{tabular}
\end{table}

The benchmark uses inner-product search. Each branch is L2-normalized independently; the compact row truncates the complementary branch to 256 coordinates and renormalizes it before concatenation. Documents and queries are both stored or queried as $[\zold;\znew]/\sqrt{2}$, so the measured blend is $\alpha=0.5$. A separate HNSW graph is built for each representation. Replacing these stored weights by another symmetric $\phi_\alpha$ requires rebuilding that representation's index; the query-only construction in Section~\ref{sec:method_alpha_kernel} avoids this rewrite but is not part of this cost benchmark. The reported recall and search work apply only to these measured configurations.

On an isolated H200, batch-one text-encoder p50 is $16.6/16.7/27.9$ms for Frozen/merged LoRA/CoSE. CoSE costs $1.67\times$ Frozen at batch one and $2.1$--$2.2\times$ at batches $32$--$64$ (QPS$_{\mathrm{Frozen}}$/QPS$_{\mathrm{CoSE}}$, using external queries). Encoder, preprocessing, and ANN costs are measured separately; full procedures and image-query results are in Appendix~\ref{app:reproducibility}.

At $\alpha=0.5$, the $[2048\|256]$ configuration obtains ROCO domain i2t R@$1$ of $0.338$ (full width $0.355$, Frozen $0.132$; three seeds). Its domain/Flickr/COCO averages are $0.342/0.883/0.675$ and ImageNet Top-1 is $0.618$, all general means within $5$pp of Frozen. These test-quality measurements and the $65$k-train-corpus ANN benchmark share a configuration, not an evaluation task. In a separate endpoint diagnostic, a full-width graph built over unweighted $[\zold;\znew]$ documents is queried with $[\zold;0]$. Its top-$10$ set overlap with a legacy graph is $98.4\%$ (Appendix~\ref{app:reproducibility}), despite equal underlying frozen scores. This is not a reuse of Table~\ref{tab:serving}'s graph; retaining the legacy index preserves the original rollback search path.

\paragraph{Storage with exact ANN rollback.}
Table~\ref{tab:serving} counts the active index. Retaining the unchanged Frozen HNSW graph additionally costs 846 MB at the projected 100k-document scale. The compact and full CoSE configurations then require 1795 MB and 2512 MB in total, respectively. This optional retained graph preserves the original approximate-search path; keeping the frozen coordinates alone preserves scores but does not force the expanded graph to visit the same candidates. Thus the 6.25--25\% vector-width overhead available with short complementary channels is neither an end-to-end latency bound nor the total storage cost of retaining an additional rollback index.

\subsection{Scale and Backbone Transfer}
\label{app:scale}
\label{sec:transfer}
\begin{table}[t]
\centering\small
\caption{ROCO test i2t R@$1$ at validation-selected $\alpha^*$, seed $0$. All rows use the same validation-selection and test-evaluation protocol. The matched-batch comparison is single-seed.}
\label{tab:scale-control}
\begin{tabular}{lcc}\toprule
Model & Training batch & R@1 \\ \midrule
CoSE 2B & 32 & 0.397 \\
CoSE 2B & 8 & 0.334 \\
CoSE 8B & 8 & 0.384 \\
\bottomrule\end{tabular}
\end{table}
At matched batch size eight, 8B exceeds 2B by $5.0$pp (Table~\ref{tab:scale-control}). Reducing the 2B batch from $32$ to eight lowers R@$1$ by $6.3$pp, consistent with batch size contributing to the apparent regression in the unmatched comparison. These single-seed comparisons neither isolate the cause nor establish universal scaling. The validation-selected weights for 2B batch 32, 2B batch eight, and 8B batch eight are $0.25$, $0.15$, and $0.25$, respectively.

On SigLIP2-so400m~\cite{SigLIP2ImprovedSigmoidLossForLanguageImage}, a variant with an additional trainable projection MLP uses $7.96$M LoRA and $5.31$M projection parameters. CoSE at $\alpha=0.5$ improves seed-$0$ i2t R@$1$ from $0.113$ to $0.321$ on ROCO and $0.059$ to $0.147$ on RSICD. On ROCO only, $\alpha=0.9$ keeps Flickr and COCO within $5$pp of Frozen; that diagnostic uses a different domain subset. RSICD general retention and ImageNet transfer were not evaluated. These results extend the evidence beyond Qwen3-VL without establishing joint attainment of those gains and preservation at $\alpha=0.5$.

\section{Analysis and Ablations}
\label{sec:analysis}

We focus the main text on two ablations that support the paper's core claims directly. Additional analyses (orthogonality geometry, $\znew$ standalone performance, rank ablation, batch-size ablation, post-hoc Frozen+LoRA score-ensemble) are in Appendix~\ref{app:ablations}.

\subsection{Joint Training versus Post-hoc Score Blending}
\label{sec:naive-concat}

A natural control blends Frozen with a separately trained LoRA encoder using the same score form and validation grid as CoSE. Domain-only tuning selects weights near the LoRA endpoint on three domains, but interior blends can improve general retention. Inference-time blending is therefore not unique to CoSE, and post-hoc blends can reach overlapping domain--general trade-off regions. On SemArt the validation-selected ensemble obtains domain R@$1$ $0.518$ versus CoSE's $0.490$, with all three seeds satisfying the general budget; the SemArt/RSICD ensemble includes retrained LoRA seeds 42/123. CoSE's joint training determines the complementary channel available throughout its operating range; the following ablation tests the role of its supervision directly.

\subsection{Loss-Term Ablation}
\label{sec:loss-ablation}

We compare HAN loss terms with three paired seeds (0/42/123), matched initialization and data order, batch 64, 8000 steps, and $\lambdaorth=0$ (Table~\ref{tab:loss-ablation}). Each configuration selects $\alpha^*$ on validation i2t R@$1$.

\begin{table}[t]
\centering\small
\caption{Paired HAN loss controls (mean$\pm$sample std, three seeds). Reference uses standalone weight 1, MRL weight 0.5, and $\lambdaorth=0$; the main-table checkpoint additionally includes $\lambdaorth=0.1$. Overlap is agreement between the two branches' top-1 retrieval results.}
\label{tab:loss-ablation}
\begin{tabular}{lccc}\toprule
Variant & Default i2t ($\alpha{=}0.5$) & $\alpha{=}0.1$ i2t & $\zold\!/\!\znew$ overlap@1 \\ \midrule
Reference & $0.420\pm0.003$ & $0.503\pm0.010$ & $25.3\pm0.2$\% \\
w/o standalone & $0.439\pm0.004$ & $0.187\pm0.004$ & $2.7\pm0.3$\% \\
w/o MRL & $0.411\pm0.002$ & $0.515\pm0.008$ & $27.5\pm0.9$\% \\
\bottomrule\end{tabular}
\end{table}

Removing standalone supervision collapses low-$\alpha$ retrieval: mean i2t R@$1$ at $\alpha=0.1$ falls from $0.503$ to $0.187$. Although midpoint i2t improves from $0.420$ to $0.439$, bidirectional domain retrieval decreases. At $\alpha=0.5$, COCO and Flickr means fall $5.6$ and $5.5$pp below Frozen; each seed exceeds the general-preservation budget. MRL remains in this control but supervises blended representations, not the standalone channel.

MRL improves HAN blended retrieval at widths $512/1024$ by $1.0$--$2.2$pp per paired seed, at a $1.6$pp mean cost to validation-selected full-width retrieval. Its benefit is specific to these truncated blends; the ROCO control does not establish the same gain (Appendix~\ref{app:truncation}).

\section{Discussion and Conclusion}
\label{sec:discussion}

\paragraph{Cost and deployment boundary.}
Exact legacy-score preservation does not require doubling vector width: retaining the full frozen channel and adding $128$--$512$ complementary coordinates gives $6.25$--$25\%$ width overhead for $d=2048$. This preserves the frozen endpoint, not necessarily full-width domain quality. Query encoding remains more expensive: full branching costs $2.1$--$2.2\times$ Frozen at saturated batches. A verified prefix-sharing implementation costs $1.58\times$ at half depth; the duplicated-batch reference does not obtain that reduction. Known-domain two-table routing has higher mixed-traffic quality in our tested settings, while CoSE offers a domain--general blend within one expanded representation.

\paragraph{Quality and scope.}
CoSE is domain-competitive rather than domain-best. Domain-optimal weights need not satisfy the general-preservation budget, and partial backfill does not provide safe blended domain gains. The matched-batch 8B comparison and SigLIP2 transfer are single-seed evidence; neither establishes universal scaling or backbone-general performance guarantees. Retaining $K$ domain branches concurrently requires $d+\sum_{k=1}^K m_k$ coordinates: truncation reduces the constant, not linear growth. Sequential updates to one domain can replace the previous branch after backfill. Concurrent multi-domain adaptation and adaptive per-query blending remain future work.

\paragraph{Conclusion.}\label{sec:conclusion}
CoSE adds domain specialization while preserving the deployed score channel within an expanded representation. Joint training supports a tunable domain--general operating range; truncation and partial branching control its costs. Domain gains require corpus backfill, with frozen service until coverage completes. CoSE is useful when retaining an auditable frozen-score endpoint and controlling general retention within one representation justify the additional compute and storage.


\clearpage
\section*{Acknowledgements}
This research was supported by the `Advanced GPU Utilization Support Program'
funded by the Government of the Republic of Korea (Ministry of Science and ICT).

\paragraph{Competing interests.}
The authors declare no additional competing interests beyond their disclosed affiliations.

\paragraph{Author Contributions}

Donghoon Han established the concept, problem definition, and logical framework of the study, and led the direction, development, and implementation of the methodology. He also led the experiments, the analysis and visualization of results, and the writing of the initial draft, contributing to every stage of the work from conception to completion.

SungHyun Moon contributed to the discussions that validated and refined the core idea and the proposed method, and to the planning of additional experiments. He also led the direction of revisions to the manuscript and the reviewer responses. 

The two authors jointly wrote the reviewer responses, took part in the author–reviewer discussion, and prepared the resubmission and camera-ready manuscripts.

SeungJae Lee reviewed the final manuscript.

\bibliographystyle{plainnat}
\bibliography{references}

\newpage
\appendix
\section{Additional Ablations}
\label{app:ablations}

This appendix extends the analyses in Section~\ref{sec:analysis}.

\subsection{Orthogonality-Penalty Diagnostics}
\label{app:orthogonality}

The main-table setting includes an output-space orthogonality penalty $\Lorth$ of weight $\lambdaorth=0.1$; the paired loss controls omit it. The penalty averages squared cross-correlations of unit-normalized channels. Its weighted contribution is bounded by $\lambdaorth/d^2$, at most $2.4\times10^{-8}$ for $d=2048$, and is therefore numerically tiny. The following single-run diagnostics and the paired controls are separate experiments.

For full-width training, let $Q_0,Q_1\in\mathbb{R}^{B\times d}$ contain the unit-normalized frozen and complementary query embeddings as rows. The penalty is
\[
 \Lorth=\frac{1}{d^2}\left\|\frac{Q_1^\top Q_0}{B}\right\|_F^2.
\]
It acts on query embeddings only, with the frozen channel detached. Since each outer product of unit vectors has Frobenius norm one, the averaged cross-correlation has Frobenius norm at most one; hence $\lambdaorth\Lorth\le\lambdaorth/d^2$.

We define overlap@$1$ as the fraction of queries for which the frozen and complementary channels retrieve the same top-ranked candidate, computed separately for i2t and t2i. It measures agreement between channels, not correctness against the ground-truth pair.

\paragraph{Diagnostic: $\lambdaorth{=}0.1$ vs $\lambdaorth{=}0$.} Comparing the reported reference checkpoint ($\lambdaorth{=}0.1$) against a separate single run with $\lambdaorth{=}0$ on $1000$ HAN extended val samples:

\begin{center}\small
\begin{tabular}{lccc}
\toprule
Metric & $\lambdaorth{=}0.1$ & $\lambdaorth{=}0$ & $\Delta$ \\
\midrule
$\cos(\zold,\znew)$ image  & $0.596 \pm 0.048$ & $0.590 \pm 0.046$ & $-0.006$ \\
$\cos(\zold,\znew)$ text   & $0.504 \pm 0.112$ & $0.492 \pm 0.124$ & $-0.012$ \\
overlap@1 i2t / t2i        & $26.0$ / $27.6\%$ & $26.9$ / $26.8\%$ & $\le 0.9$ pp \\
\bottomrule
\end{tabular}
\end{center}

These diagnostics show small observed changes in cosine and overlap when the penalty is removed; they are not pairwise statistical-equivalence tests.

\paragraph{$\lambdaorth$ sweep ($200\times$ range).} We sweep $\lambdaorth$ across five values on HAN extended test; a sixth point ($\lambdaorth{=}0$) was the diagnostic above (val-measured). These are single-run diagnostics, including an unseeded reference; they are not paired seeded controls. The paired experiment is Table~\ref{tab:loss-ablation}. For each nonzero-penalty run, $\alpha^*$ is selected by validation i2t R@$1$ on the 12-point grid; the table reports test performance at that selected weight. The selected weights for $\lambdaorth=0.01,0.1,0.5,1.0,2.0$ are $0.15,0.15,0.15,0.15,0.10$, respectively.

\begin{center}\small
\begin{tabular}{lcccccc}
\toprule
$\lambdaorth$ & $0$ (val) & $0.01$ & $0.1$ & $0.5$ & $1.0$ & $2.0$ \\
\midrule
i2t R@1 ($\alpha{=}0.5$) & $0.435$ & $0.420$ & $0.419$ & $0.418$ & $0.416$ & $0.415$ \\
i2t R@1 (test, $\alpha^{\!*}$) & --- & $0.516$ & $0.499$ & $0.511$ & $0.512$ & $0.504$ \\
\bottomrule
\end{tabular}
\end{center}

The reported nonzero-weight test runs span $0.005$ in default i2t R@$1$ and $0.017$ in tuned i2t R@$1$. The zero-weight column uses validation rather than test and should not be compared as an identical evaluation. These are descriptive diagnostics, not proof that the penalty has no effect for every checkpoint.

\paragraph{Standalone supervision and geometry.} Removing $\mathcal{L}_{\mathrm{NCE}}(r_1)$ reduces overlap@1 to $2.4\%$, whereas removing $\Lorth$ changes overlap by less than $1$pp in the reported HAN diagnostic. This supports an association between standalone retrieval supervision and the observed partial-overlap geometry; it does not establish a universal cosine equilibrium or isolate a rank-capacity mechanism.

\paragraph{Recommendation.}
Equation~\ref{eq:cose_loss} defines the contrastive objective with optional MRL (HAN weight 0.5; other-domain weight 0). The main-table setting additionally uses $\lambdaorth=0.1$, and its zero-penalty ablation remains a distinct run. Small diagnostic differences do not establish checkpoint equivalence.

\subsection{Discrete $\alpha$ sweep across four domains}
\label{app:alpha-sweep}

\begin{figure}[h]
\centering
\includegraphics[width=0.98\linewidth]{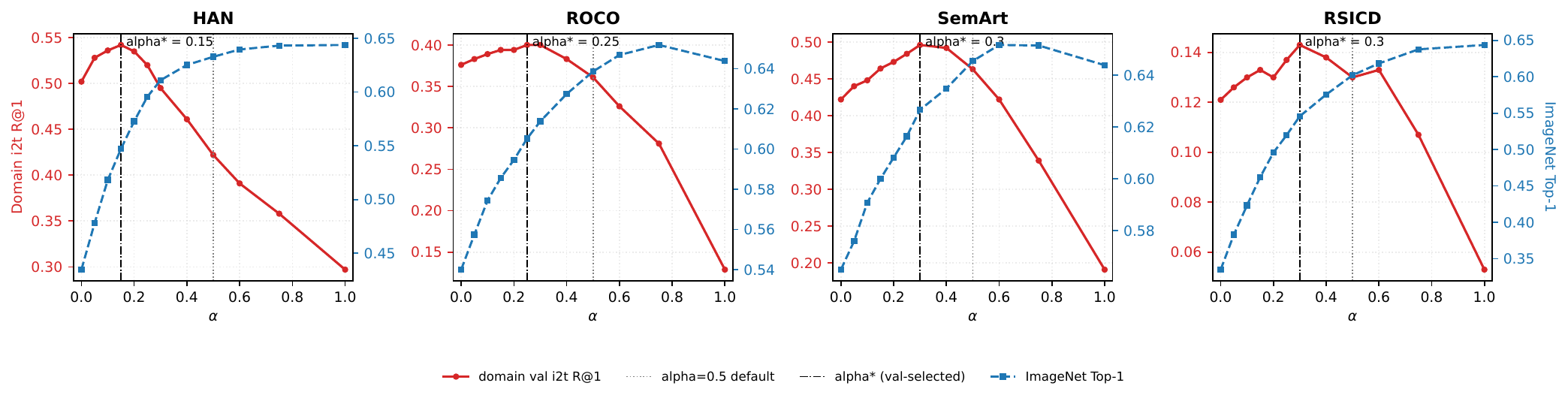}
\caption{Seed-$0$ CoSE weight sweeps across the four domains. Solid red: validation i2t R@$1$; dashed blue: ImageNet Top-1 (5000 images). Vertical guides mark $\alpha=0.5$ and validation-selected $\alpha^*$. General quality need not change monotonically, and domain-only selection need not satisfy the preservation budget.}
\label{fig:alpha-sweep}
\end{figure}

We sweep $\alpha\in\{0,0.05,0.1,0.15,0.2,0.25,0.3,0.4,0.5,0.6,0.75,1\}$, the same grid used for selection, using the main-table seed-$0$ checkpoint per domain and independent encoding instances. Figure~\ref{fig:alpha-sweep} shows validation i2t R@$1$ and ImageNet Top-1 on 5000 images. Markers use only validation i2t selection, never test results. Flickr/COCO use 1000 pairs (sampling seed 42); domain subsets use sampling seed 0.

At $\alpha=1$ all models recover the frozen score endpoint. Interior weights trade domain and general quality; other frozen-plus-adapter blends also support this sweep. Main-table means average three checkpoints, whereas these curves show seed 0 only.

\subsection{Per-domain Pareto plot}
\label{app:pareto}

Figure~\ref{fig:pareto-preview} is derived from the fully migrated, bidirectional domain and general means in Table~\ref{tab:main}. It counts benchmarks satisfying an operational $5$pp budget, not statistical significance. It makes no claim that CoSE is the only system with an exact frozen-score endpoint; Frozen+adapter blends and retained-table systems can also provide one.

\subsection{$\alpha$-swept Pareto: CoSE vs Frozen+LoRA score ensemble}
\label{app:pareto-ensemble}

Both CoSE and the Frozen+LoRA ensemble preserve the frozen score at $\alpha=1$ and require another stored channel to realize adapted retrieval. Table~\ref{tab:score-ensemble} reports fixed and domain-validation-selected ensemble operating points. Domain-only selection near the LoRA endpoint does not establish that all interior blends are redundant; preservation-constrained selection may prefer them. The loss ablation isolates the effect of standalone complementary supervision within CoSE.

\begin{table}[h]
\centering
\caption{Frozen+LoRA score ensemble vs CoSE (test split, avg(i2t,t2i) R@$1$, mean$\pm$std across 3 seeds). $\alpha^{\!*}_{\mathrm{ens}}$ is selected per seed on val i2t R@$1$ over the same $12$-value sweep used for CoSE, breaking ties toward the smallest weight. SemArt/RSICD LoRA seeds 42/123 were retrained; thus their endpoint values need not equal the original main table. Ens.\ $\alpha{=}0.5$ equals the unit-norm post-hoc concat by algebraic identity ($\cos$ of unit-concat $=\tfrac{1}{2}\cos(\zold)+\tfrac{1}{2}\cos(z_{\mathrm{lora}})$).}
\label{tab:score-ensemble}
\small
\begin{tabular}{lccccc}
\toprule
Domain & Frozen & LoRA & Ens.\ $\alpha{=}0.5$ & Ens.\ $\alpha^{\!*}_{\mathrm{ens}}$ & CoSE $\alpha^{\!*}$ \\
\midrule
HAN extended & 0.320 & 0.450 & $0.409\pm0.004$ & $0.446\pm0.010$ & $0.465\pm0.005$ \\
ROCO & 0.142 & 0.457 & $0.407\pm0.005$ & $0.458\pm0.008$ & $0.377\pm0.007$ \\
SemArt & 0.204 & 0.512 & $0.439\pm0.008$ & $0.518\pm0.007$ & $0.490\pm0.007$ \\
RSICD & 0.088 & 0.150 & $0.161\pm0.005$ & $0.162\pm0.003$ & $0.156\pm0.006$ \\
\bottomrule
\end{tabular}
\end{table}

\begin{figure}[h]
\centering
\includegraphics[width=0.95\linewidth]{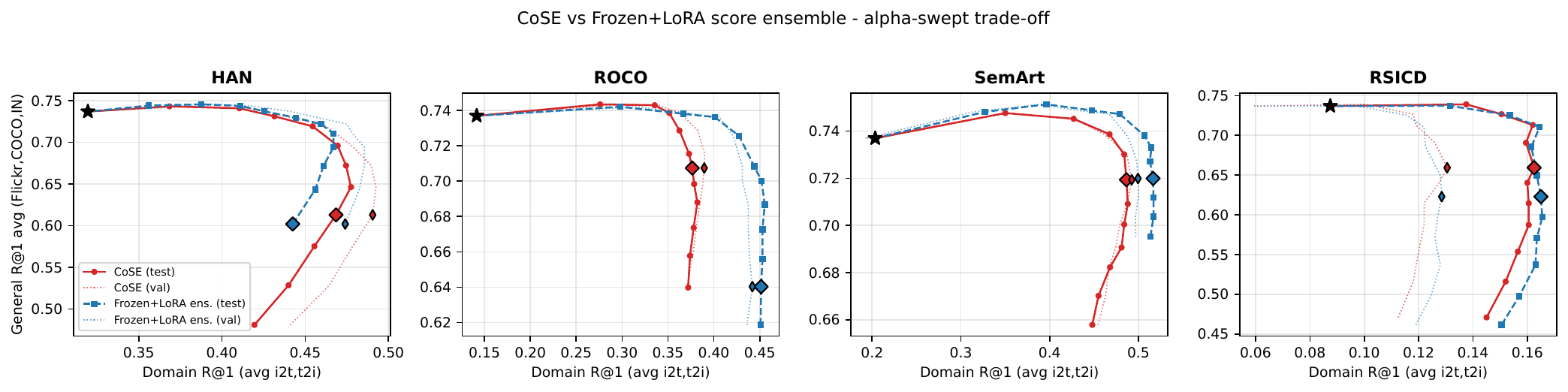}
\caption{Seed-$0$ CoSE and Frozen+LoRA trade-off curves across the four domains over the 12-point grid. Solid/dashed curves show test and dotted curves show validation domain avg(i2t,t2i) R@$1$ on the horizontal axis; the vertical axis averages Flickr/COCO bidirectional R@$1$ and ImageNet Top-1 (5000 images). Diamonds mark the same validation-i2t-selected weights on both splits; the black star marks Frozen. These are sampled trade-off curves, not a claim that selected points lie on the test Pareto frontier.}

\label{fig:pareto-ensemble}
\end{figure}

\subsection{Domain-level BC retrieval: legacy-index migration scenario}
\label{app:domain-bc}

Table~\ref{tab:main} reports fully migrated retrieval. The following cross-retrieval diagnostic uses an unchanged frozen corpus. It samples 1000 test pairs with evaluation seed equal to the training seed (0/42/123), uses FlashAttention-2, and uses the step-7500 HAN LoRA checkpoint rather than its final checkpoint. These protocol differences explain why self and Frozen references differ from Table~\ref{tab:main}; the values are not interchangeable.

\begin{table}[h]
\centering
\caption{Domain-level BC cross retrieval (avg(i2t,t2i) R@$1$, mean$\pm$sample std across 3 seeds, $1000$ pairs; variation combines training initialization and test-subset sampling). \emph{self} = method-Q $\times$ method-K (the reference reported for this diagnostic, not numerically identical to Table~\ref{tab:main}); \emph{cross} = method-Q $\times$ frozen-K (legacy-index scenario); \emph{frozen} = frozen self-retrieval anchor. \emph{\% cap} reports the fraction of self-gain over frozen recovered by cross retrieval, $(c{-}f)/(s{-}f)$. CoSE row reports the Lemma-1 prediction: BC cross against a $d$-dim frozen corpus reduces to frozen self-retrieval ranking exactly, so \% cap is $0$ (legacy users see the unchanged frozen ranking) when using the frozen channel at $\alpha=1$.}
\label{tab:domain-bc}
\scriptsize
\setlength{\tabcolsep}{4pt}
\begin{tabular}{ll cccc | c}
\toprule
Domain & Method & self & cross & frozen & \% cap & avg \% cap \\
\midrule
\multirow{4}{*}{HAN}    & LoRA          & $0.456 \pm 0.019$ & $0.390 \pm 0.008$ & $0.321$ & $50.8\%$ & \\
                         & BCT-canonical & $0.494 \pm 0.016$ & $0.424 \pm 0.003$ & $0.321$ & $59.2\%$ & \\
                         & BCT-cosine    & $0.480 \pm 0.007$ & $0.369 \pm 0.005$ & $0.321$ & $30.2\%$ & \\
                         & L2C-matched   & $0.388 \pm 0.004$ & $0.341 \pm 0.002$ & $0.321$ & $28.7\%$ & \\
\midrule
\multirow{4}{*}{ROCO}   & LoRA          & $0.452 \pm 0.011$ & $0.247 \pm 0.006$ & $0.133$ & $35.8\%$ & \\
                         & BCT-canonical & $0.369 \pm 0.010$ & $0.263 \pm 0.010$ & $0.133$ & $54.7\%$ & \\
                         & BCT-cosine    & $0.362 \pm 0.011$ & $0.205 \pm 0.006$ & $0.133$ & $31.5\%$ & \\
                         & L2C-matched   & $0.267 \pm 0.012$ & $0.171 \pm 0.010$ & $0.133$ & $28.4\%$ & \\
\midrule
\multirow{4}{*}{SemArt} & LoRA          & $0.514 \pm 0.011$ & $0.341 \pm 0.011$ & $0.203$ & $44.5\%$ & \\
                         & BCT-canonical & $0.484 \pm 0.005$ & $0.388 \pm 0.002$ & $0.203$ & $65.8\%$ & \\
                         & BCT-cosine    & $0.475 \pm 0.003$ & $0.294 \pm 0.007$ & $0.203$ & $33.4\%$ & \\
                         & L2C-matched   & $0.393 \pm 0.004$ & $0.260 \pm 0.003$ & $0.203$ & $29.7\%$ & \\
\midrule
\multirow{4}{*}{RSICD}  & LoRA          & $0.151 \pm 0.003$ & $0.108 \pm 0.008$ & $0.084$ & $35.5\%$ & \\
                         & BCT-canonical & $0.160 \pm 0.010$ & $0.132 \pm 0.009$ & $0.084$ & $63.6\%$ & \\
                         & BCT-cosine    & $0.166 \pm 0.008$ & $0.113 \pm 0.011$ & $0.084$ & $35.5\%$ & \\
                         & L2C-matched   & $0.135 \pm 0.004$ & $0.094 \pm 0.002$ & $0.084$ & $19.9\%$ & \\
\midrule
\multicolumn{6}{l}{\textbf{Per-method averages across 4 domains}} \\
\multicolumn{2}{l}{LoRA}          & & & & & $41.7\%$ \\
\multicolumn{2}{l}{BCT-canonical} & & & & & $60.8\%$ \\
\multicolumn{2}{l}{BCT-cosine}    & & & & & $32.7\%$ \\
\multicolumn{2}{l}{L2C-matched}   & & & & & $26.7\%$ \\
\multicolumn{2}{l}{\textbf{CoSE} (frozen-channel retrieval, $\alpha=1$)} & & & & & $\mathbf{0\%}$ \\
\bottomrule
\end{tabular}
\end{table}

\paragraph{Inversion of the general-BC story.}
In the separately reported general-benchmark diagnostic, LoRA cross retrieval can exceed LoRA self retrieval (RSICD-trained LoRA on Flickr: $0.733$ versus $0.643$). These values are retained as that diagnostic's references, not substituted for the differently reported means in Table~\ref{tab:main}. On domain data, Table~\ref{tab:domain-bc} reports lower cross than self retrieval. Cross and fully migrated retrieval answer different corpus-state questions.

\paragraph{CoSE's structural BC sidesteps this.}
By Lemma~\ref{lem:index_preservation}, CoSE queries using the frozen channel at $\alpha=1$ preserve legacy scores. Domain gains require the populated expanded corpus. Exact pairwise scores do not imply identical candidate sets from different ANN graphs; an unchanged legacy graph preserves the original search path.

\subsection{Effect of Detaching the Frozen Output}
\label{app:detach}

The reference detaches the pooled frozen output, not intermediate hidden states. A single-run unseeded diagnostic obtains HAN default test i2t R@$1$ $0.419$ with detach and $0.422$ without it. Since the base weights are frozen and LoRA deltas are confined to the second batch half, removing this detach does not train the frozen parameters or invalidate the score-preservation argument. It is a graph-separation diagnostic, not evidence that freezing fails without detach.

\subsection{What $\znew$ Learns Standalone}
\label{app:znew}

We evaluate seed-$0$ domain test i2t R@$1$ with the main-table checkpoints at $\alpha=1$ (Frozen), $0$ (complementary alone), and $0.5$ (blend):
\begin{center}\small
\begin{tabular}{lccc}\toprule
Domain & Frozen & $\znew$ alone & Blend, $0.5$ \\ \midrule
HAN extended & 0.298 & 0.455 & 0.415 \\
ROCO & 0.132 & 0.390 & 0.359 \\
SemArt & 0.196 & 0.416 & 0.467 \\
RSICD & 0.084 & 0.140 & 0.166 \\
\bottomrule\end{tabular}\end{center}
The complementary endpoint supplies domain signal; balancing it with the frozen channel controls general retention. These are i2t, single-seed measurements, whereas Table~\ref{tab:main} averages directions and seeds.

\subsection{LoRA Rank Ablation}
\label{app:rank}

We vary LoRA rank on HAN extended validation at $\alpha=0.5$:

\begin{center}\small
\begin{tabular}{lcc}
\toprule
Rank & Trainable params & HAN extended val i2t R@1 \\
\midrule
4   & 1.6M  & 0.418 \\
\textbf{16 (default)} & \textbf{6.4M} & \textbf{0.440} \\
32  & 12.8M & 0.448 \\
\bottomrule
\end{tabular}
\end{center}

Increasing rank from $4$ to $16$ improves HAN validation i2t R@$1$ by $0.022$; increasing it from $16$ to $32$ adds $0.008$. This within-method ablation measures sensitivity to adapter rank, rather than attributing differences between architectures to parameter count.

\subsection{Batch-Size Ablation}
\label{app:batch}

Contrastive retrieval often benefits from large batches. On HAN extended val, DDP with cross-GPU all-gather:

\begin{center}\small
\begin{tabular}{lcc}
\toprule
Config & i2t R@1 & t2i R@1 \\
\midrule
bs=64 (single GPU) & 0.440 & 0.491 \\
bs=128 (2-GPU DDP) & 0.444 & 0.482 \\
bs=256 (4-GPU DDP) & 0.438 & 0.471 \\
\bottomrule
\end{tabular}
\end{center}

Increasing batch size from 64 to 128 or 256 yields no consistent improvement in the reported HAN validation retrieval metrics.

\subsection{$\znew$ Truncation via Matryoshka}
\label{app:truncation}

We evaluate optional MRL with paired controls on HAN and ROCO. Each configuration is trained at seeds 0/42/123 with matched initialization and data order, $\lambdaorth=0$, and MRL weight either 0 or 0.5. MRL averages the four blended losses at complementary widths 256/512/1024/2048. The main-table HAN checkpoints use MRL; ROCO/SemArt/RSICD checkpoints do not.

\begin{center}\small
\begin{tabular}{lrrrr}\toprule
 & HAN MRL & HAN no MRL & ROCO MRL & ROCO no MRL \\ \midrule
Full, $0.5$ & $0.432\pm0.004$ & $0.432\pm0.004$ & $0.349\pm0.005$ & $0.360\pm0.003$ \\
Full, validation-selected $\alpha^*$ & $0.469\pm0.008$ & $0.485\pm0.009$ & $0.371\pm0.005$ & $0.387\pm0.004$ \\
$m=256$, $0.5$ & $0.420\pm0.003$ & $0.415\pm0.014$ & $0.345\pm0.004$ & $0.343\pm0.002$ \\
$m=512$, $0.5$ & $0.444\pm0.004$ & $0.430\pm0.003$ & $0.354\pm0.003$ & $0.347\pm0.004$ \\
$m=1024$, $0.5$ & $0.450\pm0.004$ & $0.433\pm0.002$ & $0.355\pm0.002$ & $0.353\pm0.007$ \\
\bottomrule\end{tabular}\end{center}
Entries are test avg(i2t,t2i) R@$1$, mean$\pm$sample std. For HAN, MRL minus no-MRL paired gains at $m=512$ are $1.6/1.0/1.6$pp and at $m=1024$ are $2.2/1.1/2.0$pp; at $m=256$ the signs differ. The full-width validation-selected mean is $1.6$pp lower with MRL. On ROCO, truncation gains are at most $0.7$pp at the midpoint, comparable to observed same-seed retraining variation, while full-width means decrease $1.0$--$1.7$pp. Thus the demonstrated MRL benefit concerns HAN truncated blends, not all domains or standalone complementary retrieval.

For the main-table ROCO checkpoint family, the three-seed $[2048\|256]$ configuration at $\alpha=0.5$ obtains domain avg R@$1$ $0.342$, Flickr $0.883$, COCO $0.675$, and ImageNet Top-1 $0.618$ (Section~\ref{sec:serving}). This result is without MRL. Truncating to $m=128$ would give $6.25\%$ width overhead but is outside the MRL training set.

\subsection{Partial Branching across all four domains}
\label{app:partial-branching-results}

Partial branching restricts adapters to layers $K$ through $L-1$. Computing the prefix once would give analytical cost $1+(L-K)/L$; the reference duplicated-batch implementation recomputes it. The quality diagnostics below concern adapter placement, not measured speed. Table below evaluates $K{=}0$ (full), $K{=}7$ ($75\%$ overhead), and $K{=}14$ ($50\%$ overhead) across all four domains (i2t R@$1$, val at $\alpha{=}0.5$).

\begin{center}\small
\setlength{\tabcolsep}{6pt}
\begin{tabular}{lcccc}
\toprule
Domain & $K=0$ & $K=7$ & $K=14$ & LoRA ref \\
\midrule
HAN extended   & 0.440 & 0.414 & 0.405 & 0.543 \\
ROCO           & 0.411 & 0.391 & 0.376 & 0.358 \\
SemArt         & 0.484 & \textbf{0.484} & 0.454 & 0.497 \\
RSICD          & 0.166 & 0.159 & 0.146 & 0.154 \\
\bottomrule
\end{tabular}
\end{center}

At $K=14$, HAN i2t R@$1$ is $0.035$ below full depth; SemArt at $K=7$ equals its full-depth diagnostic. These adapter-placement diagnostics do not imply free computation. For the $K=14$ ROCO checkpoint, prefix sharing preserves $\zold/\znew$ bit-for-bit on 200 texts and 100 images in bf16, and preserves $\alpha=0.5$ scores and rankings exactly. The frozen image output also matches an independent Frozen forward bit-for-bit. An isolated H200 text benchmark (50 warmup, 150 queries, three repeats) measures p50 $15.6/21.4/20.8$ms for Frozen/reference/prefix-sharing. At batch 64, reference and prefix-sharing cost $2.08\times/1.58\times$ Frozen. Throughput uses external query counts; the analytical $1.5\times$ is not a measured latency ratio.

\subsection{Full HAN extended Partial-Branching Sweep}
\label{app:partial-branching}

Full start-layer sweep (HAN extended val, $\alpha{=}0.5$):

\begin{center}\small
\begin{tabular}{lcccccc}
\toprule
                   & $K{=}0$ (full) & $K{=}4$ & $K{=}7$ & $K{=}14$ & $K{=}21$ & LoRA ref \\
Analytical extra cost & 100\%          & 86\%    & 75\%    & 50\%     & 25\%     & 0\% \\
Trainable params   & 6.4M           & 5.5M    & 4.8M    & 3.2M     & 1.6M     & 6.4M \\
\midrule
HAN extended i2t R@1        & 0.440 & 0.427 & 0.414 & 0.405 & 0.377 & 0.543 \\
HAN extended t2i R@1        & 0.491 & 0.453 & 0.442 & 0.431 & 0.412 & 0.400 \\
\bottomrule
\end{tabular}
\end{center}

At $K=21$, HAN validation i2t/t2i R@$1$ is $0.377/0.412$, compared with $0.440/0.491$ at full depth. The listed compute savings assume a shared-prefix implementation; the duplicated-batch reference does not realize them. Index preservation (Lemma~\ref{lem:index_preservation}) holds at every $K$.

\section{Baseline Implementation Details}
\label{app:baselines}

\paragraph{BCT --- two variants.}
We report two BCT formulations:
\begin{enumerate}
\item \textbf{Canonical (Shen et al.\ 2020)} --- the main-table version. The loss is symmetric cross-encoded InfoNCE: $\mathcal{L}_{\mathrm{bct}} = \tfrac12[\mathrm{InfoNCE}(f_{\mathrm{new}}(q), f_{\mathrm{old}}(d^+), \mathrm{negs}_{\mathrm{old}}) + \mathrm{InfoNCE}(f_{\mathrm{old}}(q), f_{\mathrm{new}}(d^+), \mathrm{negs}_{\mathrm{new}})]$. This directly targets ``new queries retrieve correctly from the old index,'' the property backward compatibility is defined by.
\item \textbf{Cosine-alignment distillation} --- a simpler feature-distillation variant: $\mathcal{L}_{\mathrm{bct}}^{\mathrm{cos}} = \mathbb{E}_x\,[1 - \cos(f_{\mathrm{new}}(x), f_{\mathrm{old}}(x))]$. This pulls new outputs toward old outputs in feature space, rather than targeting retrieval-against-old directly.
\end{enumerate}
Both BCT variants use the same $6.42$M LoRA configuration and reference schedule. CoSE batches are 64 on HAN and 32 on the other domains; matched parameter counts and schedules do not imply that every baseline uses the same batch size. The run-specific launch configurations for seed-0 BCT-canonical and all twelve L2C-matched runs specify microbatch 16 with four accumulation steps. These launch settings are not independently verified as the effective settings of every final checkpoint; accumulation does not increase the simultaneous contrastive candidate set. These are parameter-count comparisons, not comparisons with identical contrastive candidate sets across every method. The SemArt/RSICD LoRA replicas used for the ensemble and two-table comparisons use batches 32/128, respectively. Table~\ref{tab:main} uses the baseline checkpoints from the single-encoder evaluation. Canonical BCT directly supervises cross-encoded retrieval, whereas cosine alignment supervises feature proximity; neither gives CoSE's structural legacy-score equality.

\paragraph{L2C --- retrieval adaptation.}
We adapt L2C~\cite{L2CLearningToComplement} to retrieval with the following combining rule and side-branch configuration:

\emph{Element-wise reverted attention (faithful).} The official combining computes
\begin{center}
$m = \mathbf{I}_{\mathrm{CLIP}} \odot \mathbf{I}_{\mathrm{Side}}, \quad A = \mathbf{1} - \mathrm{softmax}(m, \mathrm{dim}{=}{-}1), \quad \tilde{\mathbf{I}}_{\mathrm{Side}} = \mathbf{I}_{\mathrm{Side}} \odot A$
\end{center}
where $\odot$ is element-wise multiplication and softmax is taken over the feature dimension. We use this element-wise rule rather than a matrix-product attention rule. A LayerNorm follows the gated output before combining.

\emph{Side branch reads backbone hidden states (adaptation).} The official L2C uses a fully-parallel separate encoder (SideNet) processing raw image tokens; adapting that to a decoder-based VLM like Qwen3-VL would require training a separate VLM encoder, an orthogonal design change. For comparison with CoSE's side-branch LoRA, our CPNet reads the backbone's detached final hidden states as input. The combining rule (element-wise reverted attention $+$ addition) follows the official implementation.

\emph{Parameter-matched configuration (L2C-matched).} Since the official CPNet (3 transformer blocks at backbone hidden dim, $\sim\!155$M trainable) is not parameter-comparable to our $6.4$M baselines, we report a parameter-matched variant: CPNet uses one block of width 576, eight attention heads, and an FFN width of $4\times576=2304$, with learned projections from 2048 to 576 and back, giving $\approx\!6.4$M trainable parameters. This provides a parameter-count control for comparison with CoSE. We also evaluate a full-capacity variant ($155$M trainable, three blocks, sixteen heads). The full variant uses width 2048 and FFN width 8192. The matched variant controls parameter count, not all training conditions or the isolated effect of the combining rule; neither retrieval adaptation is the original classification task.

\paragraph{Objectives and candidate sets.}
Table~\ref{tab:training-objectives} separates the implementations used for the reported model families. Standard symmetric InfoNCE averages the two directions of a $B\times B$ similarity matrix, with one diagonal positive per direction and no duplicated positive. In contrast, the directional implementation in Section~\ref{sec:method_objective} has $B+1$ logits before same-image masking, including two copies of the positive. Gradient accumulation accumulates gradients, not candidates: microbatch 16 with four accumulation steps still provides only 16 candidates per directional standard-InfoNCE query, of which 15 are negatives. Batch sizes below are paired-record counts, not the duplicated input count inside CoSE. The CoSE and SigLIP2 reference trainers backpropagate each minibatch without gradient accumulation. Parameter matching does not control the objective or contrastive candidate set and therefore does not isolate architecture alone.

\begin{table}[h]
\centering\small
\caption{Training-objective mapping. D denotes the directional duplicated-positive loss with same-image masking; S denotes standard symmetric InfoNCE. The objectives and contrastive candidate sets differ across model families.}
\label{tab:training-objectives}
\begin{tabular}{p{0.22\linewidth}p{0.69\linewidth}}
\toprule
Model family & Objective and scope \\
\midrule
Main CoSE & D on blend and complementary channel; orthogonality weight 0.1. HAN: MRL 0.5, batch 64; other domains: MRL 0, batch 32. \\
Paired CoSE controls & Same D; orthogonality 0. Standalone and MRL terms vary as specified in the ablation. \\
LoRA & HAN uses D on the sampled pair types; other domains use S. No CoSE standalone or MRL term. \\
BCT-canonical & S on adapted pairs plus 0.5 times the mean of two cross-encoded S losses (new/old and old/new). \\
BCT-cosine & S plus 0.5 times the mean query/positive cosine-alignment penalty. \\
L2C-matched/full & S on the combined embeddings. No duplicated positive or CoSE MRL term. Matched launch settings: microbatch 16, accumulation 4; effective settings are not verified for every checkpoint. \\
SigLIP2 CoSE & D on blend and complementary channel; orthogonality 0.1, no MRL; batch 32, 4000 steps, seed 0. \\
\bottomrule
\end{tabular}
\end{table}

\paragraph{SigLIP2 complementary projection.}
The SigLIP2-so400m variant applies rank-16 LoRA to all attention $Q/K/V/O$ projections in both towers. Its frozen channel uses the native pooled output. In the trainable pathway, vision hidden states are mean-pooled over tokens, while text uses the last token; each modality has a separate Linear--GELU--Linear projection before branch L2 normalization. The first linear map takes the tower hidden width to the shared embedding width, and the second preserves that width. Both towers and the embedding have width 1152, giving two $1152\to1152\to1152$ MLPs (5.31M parameters including biases). Training uses AdamW, learning rate $2\times10^{-4}$, weight decay 0.01, 200-step warmup followed by cosine decay, and temperature 0.02. This projection-equipped variant differs architecturally from the Qwen implementation.

\section{Dataset Details}
\label{app:datasets}

\paragraph{Retrieval evaluation protocol.} Each domain evaluation samples 1000 image records without replacement and selects one caption per record, using sampling seed 0 for both validation and test subsets. HAN selects one of 11 caption types uniformly, then selects English or Korean with equal probability. RSICD selects one of the available captions uniformly; ROCO and SemArt use the sole available caption. Flickr30K and COCO evaluation use sampling seed 42 and the first caption stored for each sampled image. COCO samples 1000 images from the full 5000-image validation split; ImageNet samples 5000 images from the complete validation set with seed 42. These general evaluation subsets are shared across methods. For these one-to-one retrieval evaluations, the image and selected caption from the same sampled row are the sole positive pair in either direction. Unselected captions are absent from the candidate set; this is not a multi-caption relevance evaluation.

\paragraph{HAN extended.}
HAN extended is an extended version of the HAN dataset~\cite{NIACulturalDataset2025}, currently under review for public release. We use a version with $51{,}065$ Korean-broadcast images; each image is annotated with $11$ parallel English/Korean caption pairs ($4$ factual, $6$ inferential, $1$ comprehensive; $22$ captions per image across both languages), yielding roughly $1.1$M image-caption pairs. We split deterministically $80/10/10$ (seed $42$): train $40{,}852$, val $5{,}106$, test $5{,}107$. For training, we sample four pair types: (A) image$\leftrightarrow$text ($50\%$), (B) cross-lingual EN$\leftrightarrow$KO parallel ($20\%$), (C) same-language paraphrase ($20\%$), (D) cross-lingual non-parallel ($10\%$). The base HAN dataset is publicly available on AIHub (\url{https://aihub.or.kr/aihubdata/data/view.do?dataSetSn=71866}).

\paragraph{ROCO (medical radiology).}
Radiology images with technical captions (e.g.\ ``axial CT showing hypodense lesion in segment VII of the liver''). Native train/validation/test splits (train $65{,}422$, validation $8{,}175$, test $8{,}176$).

\paragraph{SemArt (art paintings).}
$19{,}244$ paintings with long-form art-historical descriptions (artist, technique, subject). Native train/val/test splits.

\paragraph{RSICD (remote sensing).}
$\sim\!10{,}000$ satellite/aerial images with short factual captions. Native train/val/test splits. Smallest of our domains and a stress test for catastrophic forgetting (RSICD-trained LoRA obtains ImageNet Top-1 of $0.287$, seed $0$).

\paragraph{General benchmarks.}
Flickr30K~\cite{Flickr30KDataset} and COCO 2017 val~\cite{COCODataset}: $1000$ image-caption pairs each, sampled with seed $42$. ImageNet-1k~\cite{ImageNetDataset} zero-shot classification: $5000$ images, seed $42$. All external to training.

Representative image-caption samples for each domain are shown in Figs.~\ref{fig:samples-kcvu} (HAN extended), \ref{fig:samples-roco} (ROCO), \ref{fig:samples-semart} (SemArt), and \ref{fig:samples-rsicd} (RSICD).

\newcommand{\dscap}[1]{\\[2pt]{\scriptsize\itshape #1}}

\begin{figure}[t]
\centering
\begin{subfigure}[t]{0.32\linewidth}\centering
  \includegraphics[width=\linewidth]{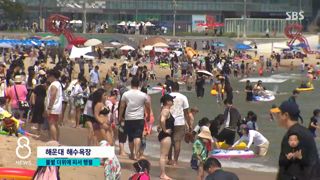}
  \dscap{``Haeundae Beach blossoms into a communal summer celebration.''}
\end{subfigure}\hfill
\begin{subfigure}[t]{0.32\linewidth}\centering
  \includegraphics[width=\linewidth]{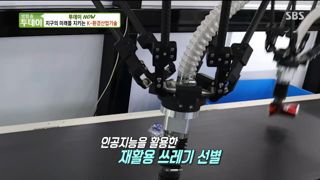}
  \dscap{``Intelligent robotic arms separate waste, signaling sustainability.''}
\end{subfigure}\hfill
\begin{subfigure}[t]{0.32\linewidth}\centering
  \includegraphics[width=\linewidth]{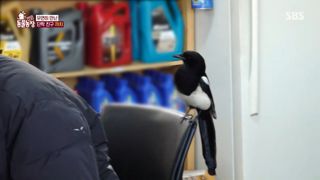}
  \dscap{``The workshop becomes a small sanctuary where dust and shared life meet.''}
\end{subfigure}
\caption{\textbf{HAN extended} (Korean broadcast, EN shown). Each image carries $11$ parallel EN/KO caption pairs; one English caption is shown per image.}
\label{fig:samples-kcvu}
\end{figure}

\begin{figure}[t]
\centering
\begin{subfigure}[t]{0.32\linewidth}\centering
  \includegraphics[width=\linewidth]{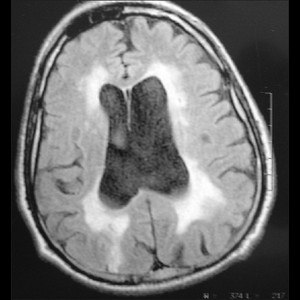}
  \dscap{``Hydrocephalus is seen on the preoperative axial MRI.''}
\end{subfigure}\hfill
\begin{subfigure}[t]{0.32\linewidth}\centering
  \includegraphics[width=\linewidth]{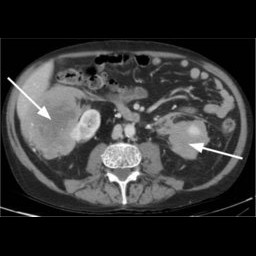}
  \dscap{``60-year-old man with relapsed multiple myeloma; CT demonstrates bilateral perirenal masses.''}
\end{subfigure}\hfill
\begin{subfigure}[t]{0.32\linewidth}\centering
  \includegraphics[width=\linewidth]{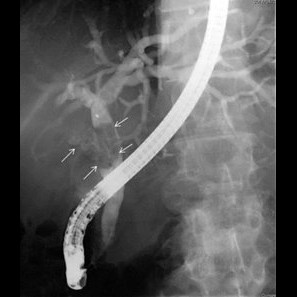}
  \dscap{``ERCP shows the mass spreading from the cystic duct to the upper-middle bile duct.''}
\end{subfigure}
\caption{\textbf{ROCO} (medical radiology). Images letterboxed to square for display; black padding preserves original content. Captions are single technical narratives.}
\label{fig:samples-roco}
\end{figure}

\begin{figure}[t]
\centering
\begin{subfigure}[t]{0.32\linewidth}\centering
  \includegraphics[height=3.4cm,keepaspectratio]{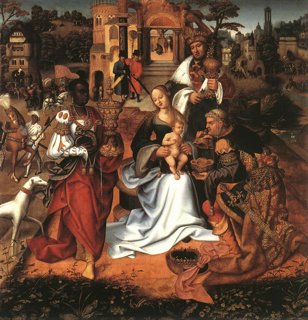}
  \dscap{``This panel is the front of the wing of an altarpiece; on the reverse is St Bernard of Clairvaux.''}
\end{subfigure}\hfill
\begin{subfigure}[t]{0.32\linewidth}\centering
  \includegraphics[height=3.4cm,keepaspectratio]{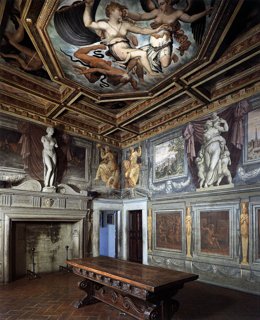}
  \dscap{``View of the Sala del Camino (Room of the Hearth), decorated by Vasari personally.''}
\end{subfigure}\hfill
\begin{subfigure}[t]{0.32\linewidth}\centering
  \includegraphics[height=3.4cm,keepaspectratio]{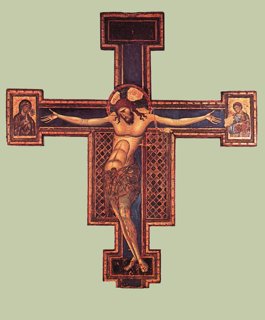}
  \dscap{``From the 13th century the Berlinghieri, a family of painters $\ldots$''}
\end{subfigure}
\caption{\textbf{SemArt} (art paintings). Portraits rendered at matched height. Captions are long-form art-historical descriptions (abbreviated here).}
\label{fig:samples-semart}
\end{figure}

\begin{figure}[t]
\centering
\begin{subfigure}[t]{0.32\linewidth}\centering
  \includegraphics[width=\linewidth]{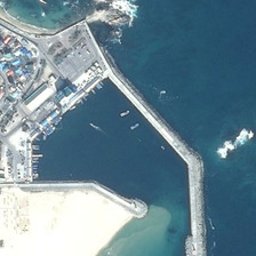}
  \dscap{``A port with several boats is close to many buildings.''}
\end{subfigure}\hfill
\begin{subfigure}[t]{0.32\linewidth}\centering
  \includegraphics[width=\linewidth]{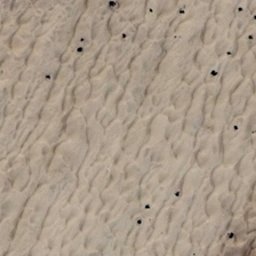}
  \dscap{``A piece of sand is like scales of fish in the desert.''}
\end{subfigure}\hfill
\begin{subfigure}[t]{0.32\linewidth}\centering
  \includegraphics[width=\linewidth]{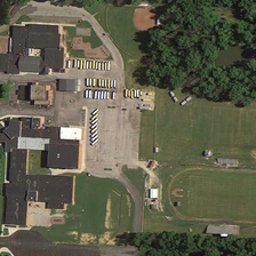}
  \dscap{``There are school buses stopping in the parking lot next to a grassland and buildings with black roof.''}
\end{subfigure}
\caption{\textbf{RSICD} (remote sensing). Square aerial tiles. Each image has five short factual captions.}
\label{fig:samples-rsicd}
\end{figure}

\section{Full R@1/R@5/R@10 i2t/t2i Breakdowns}
\label{app:preservation-detail}

This section gives the complete per-method, per-domain breakdown across all retrieval metrics: R@1, R@5, R@10 for both image-to-text (i2t) and text-to-image (t2i). Tables are split into (i) domain test (Table~\ref{tab:domain-full}), (ii) Flickr30K (Table~\ref{tab:flickr-full}), (iii) COCO 2017 val (Table~\ref{tab:coco-full}), and (iv) ImageNet-1k zero-shot (Table~\ref{tab:imagenet-full}). Entries report mean $\pm$ sample std across seeds $0$, $42$, $123$, except SemArt/RSICD LoRA general results, which use only the available original seed-$0$ checkpoints and have no std. All methods use the same general evaluation samples; domain results retain three seeds. For CoSE, two operating points are reported on domain (Table~\ref{tab:domain-full}): the zero-tune $\alpha{=}0.5$ default and the val-selected $\alpha^{\!*}$ per seed. For general retrieval (Tables~\ref{tab:flickr-full}--\ref{tab:imagenet-full}), CoSE is evaluated at $\alpha{=}0.5$ to provide a single comparable column against the non-blending baselines; Fig.~\ref{fig:alpha-sweep} shows the seed-$0$ ImageNet Top-$1$ curve over the $12$-point $\alpha$ grid.

\subsection{Domain Test (matched protocol, $1000$ pairs, seed-$0$ sampler)}

Per-method, per-domain breakdown of domain-test retrieval at all three recall cutoffs in both directions, mirroring the unified main-paper Table~\ref{tab:main} but at full granularity. Entries are mean$\pm$std across the three seeds ($0, 42, 123$). LoRA leads on i2t R@1 on HAN, ROCO, and SemArt, with a larger ROCO gap; RSICD is numerically close. On t2i, BCT-canonical leads on HAN extended at R@1, with CoSE $\alpha{=}0.5$ within $7$pp; LoRA leads on ROCO/SemArt; on RSICD the methods cluster within $\pm 2$pp.

\begin{table}[h]
\centering
\caption{Domain test --- full R@1/R@5/R@10 in both directions (i2t and t2i). Mean$\pm$sample std across $3$ seeds, all percentages. Same $1000$-pair seed-$0$ sampler used in Table~\ref{tab:main}.}
\label{tab:domain-full}
\scriptsize
\setlength{\tabcolsep}{3pt}
\begin{tabular}{ll cccccc}
\toprule
& & \multicolumn{3}{c}{i2t} & \multicolumn{3}{c}{t2i} \\
\cmidrule(lr){3-5} \cmidrule(lr){6-8}
Trained on & Method & R@1 & R@5 & R@10 & R@1 & R@5 & R@10 \\
\midrule
\multirow{6}{*}{HAN extended} & LoRA              & $52.7\pm0.5$ & $85.1\pm0.9$ & $92.9\pm0.5$ & $37.2\pm1.8$ & $69.3\pm1.1$ & $80.7\pm1.2$ \\
                              & BCT-canonical     & $47.2\pm0.8$ & $80.2\pm0.7$ & $90.1\pm0.5$ & $51.1\pm0.9$ & $81.9\pm1.0$ & $89.9\pm0.8$ \\
                              & BCT-cosine        & $45.6\pm0.9$ & $78.2\pm0.4$ & $89.0\pm0.6$ & $49.6\pm0.5$ & $81.5\pm0.3$ & $89.0\pm0.8$ \\
                              & L2C-matched       & $37.3\pm0.8$ & $68.2\pm0.7$ & $81.0\pm0.8$ & $40.4\pm0.2$ & $70.6\pm0.7$ & $81.3\pm0.4$ \\
                              & CoSE $\alpha{=}0.5$ & $41.6\pm0.4$ & $73.9\pm0.5$ & $85.6\pm0.3$ & $44.9\pm0.8$ & $76.5\pm0.6$ & $84.9\pm0.5$ \\
                              & CoSE $\alpha^{\!*}$ & $50.4\pm1.1$ & $82.6\pm0.2$ & $91.6\pm0.4$ & $42.7\pm1.3$ & $74.0\pm0.7$ & $85.0\pm1.0$ \\
\midrule
\multirow{6}{*}{ROCO}         & LoRA              & $45.6\pm0.9$ & $74.6\pm0.4$ & $84.3\pm0.2$ & $45.7\pm0.7$ & $74.2\pm0.6$ & $84.4\pm0.9$ \\
                              & BCT-canonical     & $35.9\pm0.6$ & $63.3\pm0.6$ & $75.9\pm0.7$ & $37.1\pm1.0$ & $66.5\pm0.3$ & $78.1\pm0.8$ \\
                              & BCT-cosine        & $35.5\pm0.6$ & $63.9\pm0.5$ & $76.1\pm0.8$ & $36.6\pm0.4$ & $66.9\pm1.4$ & $78.1\pm1.2$ \\
                              & L2C-matched       & $26.5\pm1.0$ & $53.8\pm0.9$ & $67.8\pm0.8$ & $28.2\pm0.4$ & $57.4\pm0.9$ & $69.9\pm0.6$ \\
                              & CoSE $\alpha{=}0.5$ & $35.5\pm0.6$ & $65.1\pm0.5$ & $77.4\pm0.6$ & $35.0\pm1.0$ & $63.8\pm0.4$ & $75.3\pm0.2$ \\
                              & CoSE $\alpha^{\!*}$ & $39.6\pm1.1$ & $69.8\pm1.0$ & $82.4\pm0.3$ & $35.8\pm0.3$ & $64.7\pm0.6$ & $75.8\pm0.4$ \\
\midrule
\multirow{6}{*}{SemArt}       & LoRA              & $51.5\pm0.4$ & $75.4\pm0.4$ & $83.5\pm0.1$ & $51.2\pm0.4$ & $76.4\pm0.7$ & $84.2\pm0.3$ \\
                              & BCT-canonical     & $47.8\pm1.0$ & $72.3\pm0.8$ & $81.7\pm1.0$ & $49.6\pm0.9$ & $74.6\pm0.6$ & $82.5\pm0.1$ \\
                              & BCT-cosine        & $45.7\pm0.9$ & $71.8\pm0.8$ & $81.0\pm0.9$ & $49.4\pm1.0$ & $72.2\pm1.1$ & $81.6\pm1.2$ \\
                              & L2C-matched       & $38.5\pm0.8$ & $63.6\pm0.5$ & $74.4\pm1.0$ & $40.0\pm1.2$ & $65.1\pm0.2$ & $75.3\pm0.6$ \\
                              & CoSE $\alpha{=}0.5$ & $47.4\pm0.6$ & $70.5\pm0.3$ & $79.7\pm0.5$ & $46.6\pm0.5$ & $70.8\pm0.9$ & $78.6\pm0.2$ \\
                              & CoSE $\alpha^{\!*}$ & $49.4\pm0.8$ & $74.7\pm1.0$ & $83.4\pm0.5$ & $48.6\pm0.6$ & $73.1\pm1.0$ & $81.5\pm0.2$ \\
\midrule
\multirow{6}{*}{RSICD}        & LoRA              & $15.6\pm0.5$ & $38.7\pm0.3$ & $53.7\pm0.4$ & $15.9\pm1.7$ & $39.5\pm0.7$ & $53.9\pm0.2$ \\
                              & BCT-canonical     & $16.0\pm1.4$ & $41.7\pm0.6$ & $58.5\pm1.2$ & $17.4\pm0.9$ & $43.4\pm0.8$ & $60.4\pm0.9$ \\
                              & BCT-cosine        & $16.5\pm1.0$ & $41.3\pm0.9$ & $58.5\pm0.8$ & $17.8\pm0.2$ & $43.6\pm1.2$ & $59.4\pm1.5$ \\
                              & L2C-matched       & $12.5\pm1.0$ & $36.2\pm0.6$ & $52.3\pm1.1$ & $14.2\pm0.5$ & $38.9\pm1.3$ & $55.3\pm0.6$ \\
                              & CoSE $\alpha{=}0.5$ & $16.5\pm0.1$ & $42.6\pm0.6$ & $58.6\pm0.3$ & $15.2\pm0.6$ & $40.5\pm0.6$ & $57.9\pm1.2$ \\
                              & CoSE $\alpha^{\!*}$ & $16.1\pm0.7$ & $41.5\pm1.2$ & $58.9\pm1.2$ & $15.0\pm0.8$ & $39.2\pm1.2$ & $56.0\pm0.6$ \\
\bottomrule
\end{tabular}
\end{table}

\subsection{Flickr30K (general retrieval, $1000$ pairs, seed-$42$ sampler)}

Flickr30K results at all six metrics (R@1/5/10 in both directions) show that CoSE at $\alpha=0.5$ remains close to Frozen. L2C-matched and BCT-cosine also retain high general retrieval quality. RSICD-trained LoRA has the largest observed drop; its general results use seed $0$ only.

\begin{table}[h]
\centering
\caption{Flickr30K full breakdown. Shared evaluation samples; mean$\pm$sample std (3 seeds), except SemArt/RSICD LoRA (seed $0$ only, no std). Frozen reference: i2t $87.4 / 98.2 / 99.6$, t2i $88.6 / 98.5 / 99.3$.}
\label{tab:flickr-full}
\scriptsize
\setlength{\tabcolsep}{3pt}
\begin{tabular}{ll cccccc}
\toprule
& & \multicolumn{3}{c}{i2t} & \multicolumn{3}{c}{t2i} \\
\cmidrule(lr){3-5} \cmidrule(lr){6-8}
Trained on & Method & R@1 & R@5 & R@10 & R@1 & R@5 & R@10 \\
\midrule
\multirow{5}{*}{HAN}   & LoRA              & $84.7\pm0.2$ & $96.6\pm0.5$ & $98.5\pm0.1$ & $70.4\pm1.4$ & $90.9\pm0.9$ & $95.4\pm0.5$ \\
                        & BCT-canonical    & $85.7\pm0.4$ & $97.3\pm0.1$ & $98.5\pm0.3$ & $85.8\pm1.2$ & $97.1\pm0.4$ & $98.6\pm0.2$ \\
                        & BCT-cosine       & $88.3\pm0.9$ & $97.9\pm0.3$ & $99.2\pm0.3$ & $89.2\pm0.5$ & $98.3\pm0.1$ & $99.0\pm0.2$ \\
                        & L2C-matched      & $85.7\pm0.7$ & $97.8\pm0.3$ & $98.9\pm0.2$ & $87.2\pm0.5$ & $97.8\pm0.2$ & $98.8\pm0.2$ \\
                        & CoSE $\alpha{=}0.5$ & $87.9\pm0.6$ & $97.9\pm0.2$ & $99.4\pm0.1$ & $88.3\pm0.2$ & $97.8\pm0.4$ & $98.9\pm0.1$ \\
\midrule
\multirow{5}{*}{ROCO}  & LoRA              & $74.6\pm0.8$ & $92.3\pm0.9$ & $95.9\pm0.5$ & $78.7\pm2.5$ & $94.4\pm0.6$ & $97.0\pm0.3$ \\
                        & BCT-canonical    & $82.7\pm0.7$ & $96.2\pm0.4$ & $98.1\pm0.3$ & $86.0\pm0.2$ & $97.0\pm0.5$ & $98.5\pm0.1$ \\
                        & BCT-cosine       & $84.4\pm0.3$ & $97.1\pm0.3$ & $98.9\pm0.2$ & $86.9\pm0.2$ & $97.5\pm0.3$ & $98.8\pm0.2$ \\
                        & L2C-matched      & $87.4\pm0.3$ & $98.2\pm0.3$ & $99.7\pm0.0$ & $89.1\pm0.2$ & $98.3\pm0.1$ & $99.1\pm0.2$ \\
                        & CoSE $\alpha{=}0.5$ & $89.0\pm0.4$ & $98.2\pm0.2$ & $99.2\pm0.1$ & $89.6\pm0.5$ & $98.8\pm0.2$ & $99.4\pm0.1$ \\
\midrule
\multirow{5}{*}{SemArt}& LoRA              & $84.7$ & $96.4$ & $98.0$ & $85.2$ & $97.3$ & $98.3$ \\
                        & BCT-canonical    & $83.1\pm1.2$ & $95.9\pm0.4$ & $98.1\pm0.2$ & $85.7\pm0.9$ & $97.4\pm0.3$ & $98.7\pm0.1$ \\
                        & BCT-cosine       & $85.5\pm0.4$ & $97.2\pm0.4$ & $98.7\pm0.2$ & $88.3\pm0.3$ & $97.9\pm0.2$ & $99.0\pm0.2$ \\
                        & L2C-matched      & $85.4\pm0.2$ & $97.7\pm0.3$ & $99.2\pm0.2$ & $87.7\pm0.4$ & $98.0\pm0.4$ & $99.1\pm0.2$ \\
                        & CoSE $\alpha{=}0.5$ & $88.1\pm0.3$ & $98.0\pm0.2$ & $99.2\pm0.2$ & $89.7\pm0.2$ & $98.8\pm0.2$ & $99.5\pm0.1$ \\
\midrule
\multirow{5}{*}{RSICD} & LoRA              & $63.8$ & $82.1$ & $87.2$ & $65.1$ & $85.3$ & $90.3$ \\
                        & BCT-canonical    & $76.3\pm1.9$ & $92.0\pm0.9$ & $95.7\pm1.1$ & $76.3\pm1.4$ & $91.9\pm1.0$ & $94.8\pm0.9$ \\
                        & BCT-cosine       & $82.3\pm2.8$ & $95.7\pm0.6$ & $97.8\pm0.7$ & $83.9\pm3.2$ & $96.4\pm1.1$ & $97.9\pm0.5$ \\
                        & L2C-matched      & $87.4\pm0.6$ & $98.2\pm0.2$ & $99.2\pm0.1$ & $87.8\pm0.2$ & $97.9\pm0.2$ & $99.2\pm0.2$ \\
                        & CoSE $\alpha{=}0.5$ & $86.0\pm0.6$ & $97.7\pm0.2$ & $98.9\pm0.3$ & $87.0\pm0.5$ & $98.2\pm0.2$ & $99.3\pm0.2$ \\
\bottomrule
\end{tabular}
\end{table}

\subsection{COCO 2017 val (general retrieval, $1000$ pairs, seed-$42$ sampler)}

Under the shared COCO evaluation, CoSE at $\alpha=0.5$ stays within $2$pp of Frozen in both R@$1$ directions. L2C-matched is within $2$pp in bidirectional R@$1$ on all four domains, and BCT-cosine within $5$pp. LoRA has larger bidirectional drops ($7.2$--$23.4$pp); SemArt/RSICD LoRA values are single-seed observations.

\begin{table}[h]
\centering
\caption{COCO 2017 val full breakdown. Shared evaluation samples; mean$\pm$sample std (3 seeds), except SemArt/RSICD LoRA (seed $0$ only, no std). Frozen reference: i2t $69.6 / 92.4 / 96.3$, t2i $67.8 / 91.3 / 96.8$.}
\label{tab:coco-full}
\scriptsize
\setlength{\tabcolsep}{3pt}
\begin{tabular}{ll cccccc}
\toprule
& & \multicolumn{3}{c}{i2t} & \multicolumn{3}{c}{t2i} \\
\cmidrule(lr){3-5} \cmidrule(lr){6-8}
Trained on & Method & R@1 & R@5 & R@10 & R@1 & R@5 & R@10 \\
\midrule
\multirow{5}{*}{HAN}   & LoRA              & $58.5\pm1.6$ & $86.1\pm0.5$ & $93.2\pm0.5$ & $42.8\pm0.9$ & $73.7\pm0.8$ & $85.4\pm0.6$ \\
                        & BCT-canonical    & $57.9\pm1.0$ & $85.1\pm0.4$ & $93.3\pm0.5$ & $56.3\pm1.3$ & $83.8\pm1.5$ & $92.5\pm1.3$ \\
                        & BCT-cosine       & $64.0\pm1.2$ & $90.1\pm0.9$ & $95.2\pm0.2$ & $65.5\pm0.6$ & $89.5\pm0.8$ & $95.2\pm0.3$ \\
                        & L2C-matched      & $69.0\pm0.4$ & $91.2\pm0.3$ & $96.3\pm0.2$ & $67.4\pm0.4$ & $90.5\pm0.2$ & $95.8\pm0.1$ \\
                        & CoSE $\alpha{=}0.5$ & $68.5\pm0.7$ & $91.8\pm0.2$ & $95.8\pm0.5$ & $68.1\pm0.2$ & $91.4\pm0.3$ & $95.7\pm0.3$ \\
\midrule
\multirow{5}{*}{ROCO}  & LoRA              & $55.7\pm1.9$ & $81.3\pm1.7$ & $89.7\pm1.1$ & $57.9\pm1.0$ & $84.0\pm1.4$ & $90.8\pm0.9$ \\
                        & BCT-canonical    & $63.8\pm0.6$ & $87.4\pm0.4$ & $93.7\pm1.3$ & $65.2\pm1.0$ & $89.3\pm0.6$ & $94.8\pm0.1$ \\
                        & BCT-cosine       & $64.0\pm1.3$ & $88.8\pm0.4$ & $95.3\pm0.2$ & $68.0\pm0.6$ & $90.3\pm0.2$ & $95.5\pm0.3$ \\
                        & L2C-matched      & $69.3\pm0.4$ & $92.1\pm0.2$ & $96.8\pm0.2$ & $67.3\pm0.3$ & $90.7\pm0.3$ & $96.6\pm0.2$ \\
                        & CoSE $\alpha{=}0.5$ & $68.9\pm0.2$ & $91.5\pm0.3$ & $96.2\pm0.2$ & $67.9\pm0.4$ & $91.1\pm0.2$ & $96.0\pm0.2$ \\
\midrule
\multirow{5}{*}{SemArt}& LoRA              & $61.2$ & $89.3$ & $95.4$ & $61.7$ & $88.8$ & $94.5$ \\
                        & BCT-canonical    & $61.3\pm0.2$ & $89.5\pm0.6$ & $94.9\pm0.6$ & $63.5\pm0.9$ & $89.1\pm0.6$ & $94.3\pm0.2$ \\
                        & BCT-cosine       & $65.7\pm1.5$ & $91.2\pm0.4$ & $96.4\pm0.3$ & $68.0\pm0.5$ & $90.5\pm0.2$ & $95.6\pm0.1$ \\
                        & L2C-matched      & $68.1\pm0.3$ & $91.3\pm0.1$ & $96.3\pm0.1$ & $65.6\pm0.3$ & $90.3\pm0.4$ & $95.8\pm0.3$ \\
                        & CoSE $\alpha{=}0.5$ & $68.3\pm0.3$ & $92.4\pm0.3$ & $96.7\pm0.3$ & $68.7\pm0.6$ & $91.4\pm0.2$ & $96.5\pm0.2$ \\
\midrule
\multirow{5}{*}{RSICD} & LoRA              & $45.2$ & $72.4$ & $82.1$ & $45.4$ & $74.2$ & $83.9$ \\
                        & BCT-canonical    & $56.5\pm0.6$ & $83.6\pm0.6$ & $91.4\pm0.5$ & $56.9\pm0.9$ & $82.9\pm1.4$ & $90.5\pm0.9$ \\
                        & BCT-cosine       & $63.8\pm3.2$ & $88.8\pm3.0$ & $94.4\pm1.5$ & $63.7\pm3.7$ & $88.2\pm1.0$ & $94.3\pm0.4$ \\
                        & L2C-matched      & $69.4\pm0.1$ & $91.8\pm0.2$ & $96.4\pm0.1$ & $66.5\pm0.1$ & $89.8\pm0.5$ & $95.6\pm0.2$ \\
                        & CoSE $\alpha{=}0.5$ & $68.0\pm0.7$ & $91.6\pm0.6$ & $95.6\pm0.5$ & $66.9\pm0.4$ & $90.0\pm0.6$ & $95.3\pm0.3$ \\
\bottomrule
\end{tabular}
\end{table}

\subsection{ImageNet-1k zero-shot ($5000$ images, seed-$42$ sampler)}

ImageNet zero-shot evaluates general visual--text alignment, with a Frozen reference of $64.4 / 86.8$ (Top-1 / Top-5). LoRA Top-1 drops range from $2.2$ to $35.7$pp; the SemArt/RSICD values use seed $0$ only. CoSE at $\alpha=0.5$ stays within $4.8$pp of Frozen on both metrics. L2C-matched also preserves ImageNet performance, with Top-1 drops of $0.7$--$1.9$pp, at lower domain quality than CoSE in Table~\ref{tab:main}.

\begin{table}[h]
\centering
\caption{ImageNet zero-shot Top-1 / Top-5. Shared evaluation samples; mean$\pm$sample std (3 seeds), except SemArt/RSICD LoRA (seed $0$ only, no std). Frozen reference: $64.4 / 86.8$.}
\label{tab:imagenet-full}
\scriptsize
\setlength{\tabcolsep}{6pt}
\begin{tabular}{ll cc}
\toprule
Trained on & Method & Top-1 & Top-5 \\
\midrule
\multirow{5}{*}{HAN}   & LoRA           & $52.4\pm1.2$ & $80.2\pm0.7$ \\
                        & BCT-canonical  & $54.2\pm2.5$ & $82.0\pm0.9$ \\
                        & BCT-cosine     & $62.3\pm1.6$ & $86.3\pm1.0$ \\
                        & L2C-matched    & $62.5\pm0.6$ & $85.7\pm0.6$ \\
                        & CoSE $\alpha{=}0.5$ & $63.5\pm0.3$ & $86.5\pm0.1$ \\
\midrule
\multirow{5}{*}{ROCO}  & LoRA           & $49.1\pm1.4$ & $73.4\pm1.3$ \\
                        & BCT-canonical  & $59.3\pm2.2$ & $82.3\pm1.8$ \\
                        & BCT-cosine     & $62.2\pm1.5$ & $85.0\pm1.1$ \\
                        & L2C-matched    & $63.7\pm0.1$ & $86.8\pm0.1$ \\
                        & CoSE $\alpha{=}0.5$ & $63.8\pm0.1$ & $86.4\pm0.1$ \\
\midrule
\multirow{5}{*}{SemArt}& LoRA           & $62.2$ & $84.7$ \\
                        & BCT-canonical  & $60.2\pm0.6$ & $83.4\pm0.7$ \\
                        & BCT-cosine     & $63.0\pm1.1$ & $85.3\pm0.7$ \\
                        & L2C-matched    & $63.3\pm0.1$ & $86.0\pm0.1$ \\
                        & CoSE $\alpha{=}0.5$ & $65.0\pm0.4$ & $87.2\pm0.1$ \\
\midrule
\multirow{5}{*}{RSICD} & LoRA           & $28.7$ & $54.2$ \\
                        & BCT-canonical  & $44.9\pm2.0$ & $71.1\pm1.4$ \\
                        & BCT-cosine     & $51.1\pm4.8$ & $77.6\pm4.0$ \\
                        & L2C-matched    & $63.1\pm0.3$ & $85.7\pm0.2$ \\
                        & CoSE $\alpha{=}0.5$ & $59.6\pm0.7$ & $83.4\pm0.7$ \\
\bottomrule
\end{tabular}
\end{table}

\section{Feature Drift: CoSE vs BCT vs LoRA}
\label{app:drift}

\noindent\textbf{Setup.} For each method and domain, we measure the cosine similarity between the method's output embedding and the frozen base encoder's embedding on the same input, averaged over $200$ val-split images (Table~\ref{tab:drift}). For CoSE we measure the $\zold$ slice (first $d$ coordinates of $\zfinal$); for LoRA / BCT-canonical / BCT-cosine we measure the full post-adapter output. CoSE's structural-preservation claim implies zero drift up to kernel determinism; BCT's distillation target implies small-but-nonzero drift; LoRA has no alignment loss and is expected to drift substantially.

\begin{table}[h]
\centering
\small
\caption{Per-domain output-embedding drift relative to the frozen base. Drift $= 1 - \cos(f_{\text{method}}(x), f_{\text{frozen}}(x))$, averaged over $200$ val images per domain, seed $0$. CoSE mean drift rounds to $0$ under bf16/SDPA independent forwards; sample std is approximately $7\times10^{-8}$. Cosine arithmetic is subject to floating-point rounding.}
\label{tab:drift}
\setlength{\tabcolsep}{6pt}
\begin{tabular}{lcccc}
\toprule
Domain & CoSE (ours) & LoRA & BCT-canonical & BCT-cosine \\
\midrule
HAN extended & $\mathbf{0.00 \times 10^0}$ & $3.58 \times 10^{-1}$ & $4.18 \times 10^{-1}$ & $9.72 \times 10^{-2}$ \\
ROCO         & $\mathbf{0.00 \times 10^0}$ & $3.03 \times 10^{-1}$ & $5.11 \times 10^{-1}$ & $6.14 \times 10^{-2}$ \\
SemArt       & $\mathbf{0.00 \times 10^0}$ & $4.47 \times 10^{-1}$ & $5.30 \times 10^{-1}$ & $9.31 \times 10^{-2}$ \\
RSICD        & $\mathbf{0.00 \times 10^0}$ & $7.23 \times 10^{-1}$ & $5.82 \times 10^{-1}$ & $1.39 \times 10^{-1}$ \\
\bottomrule
\end{tabular}
\end{table}

\paragraph{Takeaways.}
CoSE's reported mean cosine drift rounds to zero under bf16/SDPA independent forwards; this is not an FP32 experiment or a maximum-coordinate-error measurement. BCT-cosine, which explicitly minimizes $1 - \cos(f_{\text{new}}, f_{\text{frozen}})$ during training, drifts least among the adapted baselines ($0.06$--$0.14$) but is still many orders of magnitude larger than the observed cosine-rounding scale. BCT-canonical (symmetric cross-encoded InfoNCE, Shen et al.\ 2020) does not enforce feature-space proximity directly, and its drift is actually larger than LoRA on three of four domains ($0.42$--$0.58$ vs LoRA's $0.30$--$0.45$); canonical BCT targets retrieval-via-old-index consistency, not embedding proximity. LoRA drifts $0.30$--$0.72$ with no alignment target --- particularly severe on small-data RSICD where it catastrophically moves away from the frozen output. The quantitative gap between ``structural preservation'' (CoSE, $0$) and ``approximate preservation'' (BCT-cosine, $\sim 10^{-1}$) is what the paper refers to as a guarantee by construction versus a distillation approximation.

\section{Deployment Scenarios}
\label{app:deployment}

CoSE preserves the old coordinates while computing a new complementary channel. Domain gains require backfilling that channel; uninterrupted frozen service during backfill can also be provided by blue-green migration with other methods.

\paragraph{Frozen service and rollback.}
At $\alpha=1$, the complementary branch can be bypassed and queries can use the original legacy index. Scores are preserved up to execution-level numerical determinism. Switching the weight to one on a different ANN graph does not guarantee identical candidates; retaining the original graph preserves the original search path and adds storage during the rollback window. Without storage sharing, adding the projected 846MB legacy index to the compact/full expanded indices in Table~\ref{tab:serving} gives 1795/2512MB at 100k documents. These sums illustrate additional rollback storage; they are not separate measurements.

\paragraph{Backfill followed by an atomic switch.}
A rolling job populates $\znew$ while serving continues with the frozen pathway. Mixed scoring of covered and uncovered documents introduces coverage bias. Ten coverage draws per operating point yielded the following diagnostic at $25\%$ coverage; the baselines are specific to that experiment, not Table~\ref{tab:main}'s sampler references.

\begin{center}\small
\begin{tabular}{lrrrr}\toprule
 & HAN & ROCO & SemArt & RSICD \\ \midrule
Frozen R@1 & 0.341 & 0.151 & 0.211 & 0.091 \\
R@1 at $25\%$ coverage & 0.270 & 0.134 & 0.215 & 0.084 \\
Excess covered top-1 share & 0.52 & 0.71 & 0.44 & 0.65 \\
\bottomrule\end{tabular}
\end{center}
Excess share is $P(\mathrm{top\mbox{-}1}\in\mathrm{covered})-0.25$. Rank fusion and isotonic calibration reduced but did not eliminate this bias. We recommend frozen service until coverage is complete, then an atomic switch to a common selected weight.

\paragraph{Steady-state cost.}
Full-width storage is $d+m$ coordinates per document, or $d+m'$ when only the complementary channel is truncated. This preserves frozen scores, not necessarily full-width adapted quality. Query computation is approximately $1+(L-K)/L$ base forwards with a shared prefix of $K$ layers. Computing complementary coordinates for an existing document can require recomputing hidden states unless they were cached; reusing its stored final frozen vector does not by itself supply those hidden states. ANN search work is measured separately in Table~\ref{tab:serving}.

\section{Implementation and Reproducibility}
\label{app:reproducibility}

The main checkpoints use PyTorch and bf16. The domain/general, serving, and paired-control evaluations use the default SDPA backend; the separate cross-retrieval diagnostic explicitly uses FlashAttention-2. AdamW uses lr $2\times10^{-4}$, weight decay $0.01$, 200-step warmup and cosine decay through 8000 steps. Split seed is 42; domain evaluation sampling seed is 0 and general sampling seed is 42, except the explicitly labeled cross-retrieval diagnostic. Final-step checkpoints are used for the main evaluation.

\paragraph{Contrastive implementation.} The loss in Section~\ref{sec:method_objective} is directional. A positive similarity is appended to the in-batch logits, whose diagonal is also retained; hence its denominator contains two copies of the positive exponential. Only the appended logit is the cross-entropy target. Off-diagonal pairs with the same image ID are masked. This applies to the blended, standalone, and MRL losses of the reported CoSE checkpoints. The duplicated-positive objective has a different gradient from standard InfoNCE.

\paragraph{Training configurations.} The main-table CoSE models use batch 64 and MRL weight 0.5 on HAN, and batch 32 without MRL on ROCO/SemArt/RSICD; all include orthogonality weight 0.1. HAN paired loss controls and ROCO paired MRL controls use orthogonality weight 0. For the ensemble and two-table comparisons, SemArt/RSICD LoRA seeds 42/123 use retrained replicas; these are distinct from the corresponding single-encoder baseline checkpoints in Table~\ref{tab:main}.

\paragraph{Serving protocol.} Isolated H200 encoder measurements use fresh system instances, bf16/SDPA, no-grad evaluation, external query counts, 100 warmup queries, 300 batch-one queries, and three repeats. Text p50/p95 is 16.6/17.3ms for Frozen, 16.7/17.6ms for merged LoRA, and 27.9/29.1ms for CoSE. Image p50/p95 is 31.0/37.1, 29.8/33.3, and 42.5/52.2ms, respectively. CPU tokenization takes 3.7--4.0ms for text and 8.9--9.8ms for images. At text batch 64, QPS is 985 for Frozen and approximately 446 for CoSE (2.21 times cost); peak memory is 5.4GB for Frozen/merged LoRA and 6.6GB for CoSE. The ANN benchmark uses one CPU thread, 65k ROCO train images, 1000 test text queries, HNSW $M=32$, efConstruction 200, and efSearch 64/64/32/64 for Frozen/LoRA/CoSE full/compact. Recall@10 against each representation's exact search is 0.953/0.982/0.956/0.981 and search p50 is 0.357/0.469/0.414/0.418ms. These are separate encoder and search measurements, not summed end-to-end quantiles.

\paragraph{Frozen-endpoint ANN diagnostic.} Separately from Table~\ref{tab:serving}, we build an inner-product HNSW graph over 65k ROCO documents stored as $[\zold;\znew]\in\mathbb{R}^{4096}$, with each branch L2-normalized and no subsequent concatenation normalization. This graph is queried with $[\zold;0]$ for 1000 text queries, without rebuilding it at $\alpha=1$. A separately built legacy graph stores and queries only the 2048-dimensional $\zold$. Both use $M=32$, efConstruction 200, and efSearch 256. The mean query-wise set intersection $|I_{\mathrm{expanded}}^{10}(q)\cap I_{\mathrm{legacy}}^{10}(q)|/10$ is $0.984$; this is set overlap, not the fraction of identical lists. Recall against exact frozen top-$10$ is $0.982$ for expanded and $0.989$ for legacy. The expanded graph is built specifically for this diagnostic, not reused from the $\alpha=0.5$ cost benchmark, and no full-$\alpha$ cost or recall sweep is implied.

\paragraph{Conditional seed variation.} ROCO/SemArt/RSICD training uses a fixed batch sampler (seed 42), while training seeds 0/42/123 vary LoRA initialization and torch RNG. Their sample std excludes data-order variation and does not estimate kernel nondeterminism from repeated runs. ROCO same-seed retraining comparisons moved midpoint R@$1$ by 0.5--0.8pp; these two pairs differ in a numerically tiny penalty and do not isolate the source of variation. The corresponding HAN data-order sensitivity and repeated-run distribution have not been established.

\paragraph{Numerical preservation.} The score-preservation lemma is structural. The independent-instance 2B bf16/SDPA cosine diagnostic uses 200 images per domain and has sample variation of approximately $10^{-7}$. Separately, the $K=14$ shared-prefix ROCO checkpoint matches both the reference and independent Frozen image outputs bit-for-bit on 100 images. Neither observation is a universal FP32 claim or a guarantee of equal candidates from distinct ANN graphs.

\subsection{Measured Training Memory}
\label{app:memory}
We measure training memory with the actual trainer in train mode, bf16, and gradient checkpointing enabled. GPU-wide peak usage from \texttt{nvidia-smi} includes reserved memory.
\begin{center}\small
\begin{tabular}{lrr}\toprule
Configuration & Batch & Peak (GB) \\ \midrule
HAN, paired reference (MRL + standalone) & 64 & 78.9 \\
HAN, without standalone or without MRL & 64 & 70.0 \\
ROCO, MRL on/off control & 32 & 35.3 \\
\bottomrule\end{tabular}\end{center}
These GPU-wide usage peaks apply only to the listed configurations and are not allocator-only peaks. Smaller microbatches were not evaluated. Gradient accumulation does not reproduce a larger in-batch contrastive negative set.

\section{Supporting Compatibility Results}
\label{app:compatibility-theory}

The main text gives the score definition and architectural guarantee. Here we record the supporting linear-algebra statements and proofs.

\begin{theorem}[Same-channel exact compatibility is trivial under full-rank legacy indices]
\label{thm:same_channel_impossibility}
If $g:\mathcal{X}\to\mathbb{R}^d$ satisfies $Z_{\mathrm{old}}g(q)=Z_{\mathrm{old}}\zold(q)$ for every
query $q$, then
\[
    g(q)-\zold(q)\in\operatorname{Null}(Z_{\mathrm{old}}).
\]
In particular, if $\operatorname{rank}(Z_{\mathrm{old}})=d$, then $g(q)=\zold(q)$ for every
$q$.
\end{theorem}

\begin{proof}
Subtracting the two score vectors gives
$Z_{\mathrm{old}}(g(q)-\zold(q))=0$. Hence the offset lies in $\operatorname{Null}(Z_{\mathrm{old}})$. Full column rank gives $\operatorname{Null}(Z_{\mathrm{old}})=\{0\}$.
\end{proof}

\begin{theorem}[Direct-sum normal form for auditable exact compatibility]
\label{thm:direct_sum_normal_form}
Let $F:\mathcal{X}\to\mathbb{R}^{d+m}$ be an expanded representation. Suppose there exists a fixed rank-$d$ linear readout
$P:\mathbb{R}^{d+m}\to\mathbb{R}^d$ such that
\[
    PF(x)=\zold(x) \qquad \forall x\in\mathcal{X}.
\]
Then, up to an invertible linear change of coordinates,
\[
    F(x)=
    \begin{bmatrix}
    \zold(x)\\
    h(x)
    \end{bmatrix}
\]
for some $h:\mathcal{X}\to\mathbb{R}^m$.
\end{theorem}

\begin{proof}
Since $P$ has rank $d$, the rows of $P$ can be completed to a basis of the
dual space. Hence there exists $R:\mathbb{R}^{d+m}\to\mathbb{R}^m$ such that
$T(v)=[Pv;Rv]$ is invertible. Applying $T$ gives
$TF(x)=[PF(x);RF(x)]=[\zold(x);h(x)]$ with $h(x)=RF(x)$.
\end{proof}

\begin{proposition}[CoSE defines a positive-semidefinite retrieval kernel family]
\label{prop:psd_kernel}
For every $\alpha\in[0,1]$,
\[
    k_\alpha(x,y)
    =
    \alpha\langle \zold(x),\zold(y)\rangle
    +(1-\alpha)\langle \znew(x),\znew(y)\rangle
\]
is positive semidefinite and equals the inner product induced by
\[
    \phi_\alpha(x)=
    \begin{bmatrix}
    \sqrt{\alpha}\,\zold(x)\\
    \sqrt{1-\alpha}\,\znew(x)
    \end{bmatrix},
    \qquad
    k_\alpha(x,y)=\langle \phi_\alpha(x),\phi_\alpha(y)\rangle .
\]
At $\alpha=1$, $k_1$ is the frozen encoder kernel exactly.
\end{proposition}

\begin{proof}
Inner-product kernels are positive semidefinite, and nonnegative weighted sums
preserve positive semidefiniteness. The representation by $\phi_\alpha$ follows
by direct expansion.
\end{proof}

\newpage
\section*{NeurIPS Paper Checklist}
\begin{enumerate}

\item {\bf Claims}
    \item[] Question: Do the main claims made in the abstract and introduction accurately reflect the paper's contributions and scope?
    \item[] Answer: \answerYes{}
    \item[] Justification: The abstract and introduction distinguish exact legacy-score preservation at the frozen endpoint from domain quality and general retention at other weights. Table~\ref{tab:main} and Sections~\ref{sec:deployment-comparison}--\ref{sec:transfer} report operating points, deployment alternatives, serving cost, and single-seed controls. No unique dominance or safe mixed-coverage gain is claimed.

\item {\bf Limitations}
    \item[] Question: Does the paper discuss the limitations of the work performed by the authors?
    \item[] Answer: \answerYes{}
    \item[] Justification: Section~\ref{sec:discussion} covers encoding and storage costs, domain-quality gaps, partial-backfill bias, ANN rollback scope, single-seed scale/transfer evidence, and linear storage growth for concurrent branches. Dataset-release limitations are stated in the setup.

\item {\bf Theory assumptions and proofs}
    \item[] Question: For each theoretical result, does the paper provide the full set of assumptions and a complete (and correct) proof?
    \item[] Answer: \answerYes{}
    \item[] Justification: The main text states the score definition and index-preservation lemma with its architectural argument. Appendix~\ref{app:compatibility-theory} gives the supporting same-channel, direct-sum, and kernel statements and proofs with explicit assumptions. Score equality is distinguished from approximate-search agreement.

    \item {\bf Experimental result reproducibility}
    \item[] Question: Does the paper fully disclose all the information needed to reproduce the main experimental results of the paper to the extent that it affects the main claims and/or conclusions of the paper (regardless of whether the code and data are provided or not)?
    \item[] Answer: \answerNo{}
    \item[] Justification: The manuscript describes the core architecture, objective, reference penalty setting, datasets, and evaluation rules. Checkpoint-to-objective mapping and diagnostic sampling differences are specified in Appendix~\ref{app:baselines} and the implementation section. Public reproducibility is limited by the absence of an accessible checkpoint package and by restrictions on the HAN extended captions.

\item {\bf Open access to data and code}
    \item[] Question: Does the paper provide open access to the data and code, with sufficient instructions to faithfully reproduce the main experimental results, as described in supplemental material?
    \item[] Answer: \answerNo{}
    \item[] Justification: The paper does not provide a public training/evaluation-code or adapter-checkpoint access URL. Public benchmarks are cited; HAN extended captions are subject to provider review.

\item {\bf Experimental setting/details}
    \item[] Question: Does the paper specify all the training and test details (e.g., data splits, hyperparameters, how they were chosen, type of optimizer) necessary to understand the results?
    \item[] Answer: \answerNo{}
    \item[] Justification: The main text reports the reference schedule, validation selection, and primary samplers. The appendix specifies actual CoSE batch sizes and MRL usage, validation selection, diagnostic sampler differences, paired controls, and retrained baseline settings. The paper does not fully specify all baseline configurations or provide public reproduction access.

\item {\bf Experiment statistical significance}
    \item[] Question: Does the paper report error bars suitably and correctly defined or other appropriate information about the statistical significance of the experiments?
    \item[] Answer: \answerYes{}
    \item[] Justification: The main domain table reports mean and sample standard deviation across three training seeds; the domain and general appendices use the same sample-standard-deviation convention. SemArt/RSICD LoRA general results are explicitly reported for seed 0 only. Loss controls are paired across three seeds. Non-HAN CoSE seed variation is conditional on fixed data order; observed same-seed retraining variation is reported separately. L2C-full and scale/backbone controls are labeled single-seed. Preservation budgets and grayscale bins are descriptive thresholds, not pairwise significance or equivalence tests.

\item {\bf Experiments compute resources}
    \item[] Question: For each experiment, does the paper provide sufficient information on the computer resources (type of compute workers, memory, time of execution) needed to reproduce the experiments?
    \item[] Answer: \answerNo{}
    \item[] Justification: The reference trials use an H200 with the schedule in the setup; an 8000-step 2B run takes approximately six hours. Appendix~\ref{app:memory} reports actual train-mode peaks of 78.9 GB for HAN reference batch 64 and 35.3 GB for ROCO batch 32. Isolated encoding benchmarks distinguish batch-one latency, throughput, and prefix sharing. Smaller-GPU trainability and complete project-compute accounting are not provided.
    
\item {\bf Code of ethics}
    \item[] Question: Does the research conducted in the paper conform, in every respect, with the NeurIPS Code of Ethics \url{https://neurips.cc/public/EthicsGuidelines}?
    \item[] Answer: \answerYes{}
    \item[] Justification: The work uses publicly available pretrained models (Qwen3-VL-Embedding) and standard benchmarks; HAN extended is licensed via AIHub. No human-subjects research, no scraped private data, and no models with foreseeable misuse risk beyond the standard image-text retrieval setting.

\item {\bf Broader impacts}
    \item[] Question: Does the paper discuss both potential positive societal impacts and negative societal impacts of the work performed?
    \item[] Answer: \answerNo{}
    \item[] Justification: Exact frozen-score preservation can reduce behavioral uncertainty during retrieval migration, but domain adaptation still requires corpus backfill and compute. Retrieval systems can inherit biases and expose sensitive content; preserving frozen scores does not remove those risks. A dedicated discussion of societal impacts and mitigations is not provided.
    
\item {\bf Safeguards}
    \item[] Question: Does the paper describe safeguards that have been put in place for responsible release of data or models that have a high risk for misuse (e.g., pre-trained language models, image generators, or scraped datasets)?
    \item[] Answer: \answerNA{}
    \item[] Justification: We release no new pretrained generative models or scraped datasets that pose misuse risk. CoSE adapters, when released, are domain-specific embedding modules that inherit the base encoder's risk profile unchanged.

\item {\bf Licenses for existing assets}
    \item[] Question: Are the creators or original owners of assets (e.g., code, data, models), used in the paper, properly credited and are the license and terms of use explicitly mentioned and properly respected?
    \item[] Answer: \answerNo{}
    \item[] Justification: Creators and datasets are cited, including Qwen3-VL and SigLIP2. A complete per-asset license/version table and verification of all release terms are not included; HAN extended-caption release remains separately restricted.

\item {\bf New assets}
    \item[] Question: Are new assets introduced in the paper well documented and is the documentation provided alongside the assets?
    \item[] Answer: \answerNA{}
    \item[] Justification: The paper does not provide a public access path for new code or adapter checkpoints; HAN extended captions remain subject to provider review.

\item {\bf Crowdsourcing and research with human subjects}
    \item[] Question: For crowdsourcing experiments and research with human subjects, does the paper include the full text of instructions given to participants and screenshots, if applicable, as well as details about compensation (if any)?
    \item[] Answer: \answerNA{}
    \item[] Justification: No crowdsourcing or human-subjects research is conducted. All training and evaluation data come from existing public or licensed datasets.

\item {\bf Institutional review board (IRB) approvals or equivalent for research with human subjects}
    \item[] Question: Does the paper describe potential risks incurred by study participants, whether such risks were disclosed to the subjects, and whether Institutional Review Board (IRB) approvals (or an equivalent approval/review based on the requirements of your country or institution) were obtained?
    \item[] Answer: \answerNA{}
    \item[] Justification: No human-subjects research is conducted in this work. The HAN dataset's underlying captions were collected by a third-party data provider under their own consent procedures; we use the released data under that platform's license.

\item {\bf Declaration of LLM usage}
    \item[] Question: Does the paper describe the usage of LLMs if it is an important, original, or non-standard component of the core methods in this research? Note that if the LLM is used only for writing, editing, or formatting purposes and does \emph{not} impact the core methodology, scientific rigor, or originality of the research, declaration is not required.
    \item[] Answer: \answerNA{}
    \item[] Justification: LLMs are not used as a methodologically important, original, or non-standard component of the core method. Per NeurIPS policy, declaration is not required for writing/editing assistance.

\end{enumerate}

\end{document}